\documentclass{aa} 
\usepackage{graphicx}
\usepackage{url}
\usepackage[breaklinks=true]{hyperref}
\hypersetup{colorlinks=true,linkcolor=blue,citecolor=blue,filecolor=blue,urlcolor=blue}
\usepackage{twoopt}
\usepackage{natbib}
\bibpunct{(}{)}{;}{a}{}{,} %% natbib format for A&A and ApJ
\usepackage{amssymb}
\usepackage{float}
\usepackage{color}
\usepackage{epstopdf}
\usepackage{ctable}
\usepackage{multirow}
\usepackage[font=small]{caption}
\usepackage{multirow}
\usepackage{comment}
\usepackage{txfonts}

\usepackage{xcolor}

\newcommand{\jybeam}{${\rm Jy~beam}^{-1}$}
\newcommand{\rms}{{\sigma_{\rm rms}}}
\newcommand{\cluster}{PSZ2G091}

\begin{document}

\title{The LOFAR sub-arcsecond view of the high-redshift radio relic in PSZ2\,G091.83+26.11}
%\subtitle{}

\author{
G.~Di Gennaro\inst{\ref{inst:ira}}
\and
R.~Timmerman\inst{\ref{inst:durham1},\ref{inst:durham2}}
\and
M.~Hoeft\inst{\ref{inst:tauten}}
\and
F.~de Gasperin\inst{\ref{inst:ira}}
\and
R.~J.~van Weeren\inst{\ref{inst:leiden}}
\and 
A.~Botteon\inst{\ref{inst:ira}}
\and
M.~Br\"uggen\inst{\ref{inst:hamb}}
\and
J.~M.~G.~H.~J.~de Jong\inst{\ref{inst:leiden},\ref{inst:astron}}
\and 
T.~W.~Shimwell\inst{\ref{inst:astron},\ref{inst:leiden}}
\and
F.~Sweijen\inst{\ref{inst:durham1},\ref{inst:durham2}}
\and
G.~Brunetti\inst{\ref{inst:ira}}
\and
R.~Cassano\inst{\ref{inst:ira}}
\and
E.~De Rubeis\inst{\ref{inst:hamb},\ref{inst:ira}}
\and
W.~Forman\inst{\ref{inst:cfa}}
\and
H.~J.~A.~R\"ottgering\inst{\ref{inst:leiden}}
\and
A.~Simionescu\inst{\ref{inst:sron},\ref{inst:leiden}}
\and
H.~Ye\inst{\ref{inst:cambridge}}
}

\institute{
{INAF - Istituto di Radioastronomia, via P. Gobetti 101, 40129 Bologna, Italy}\label{inst:ira}\\
             \email{g.digennaro@ira.inaf.it}
\and
{Centre for Extragalactic Astronomy, Department of Physics, Durham University, Durham, DH1 3LE, UK}\label{inst:durham1}
\and
{Institute for Computational Cosmology, Department of Physics, Durham University, South Road, Durham DH1 3LE, UK}\label{inst:durham2}
\and
{Th\"uringer Landessternwarte, Sternwarte 5, 07778 Tautenburg, Germany}\label{inst:tauten}
\and
{Leiden Observatory, Leiden University, PO Box 9513, 2300 RA Leiden, The Netherlands}\label{inst:leiden}
\and
{Hamburger Sternwarte, Universit\"at Hamburg, Gojenbergsweg 112, 21029 Hamburg, Germany}\label{inst:hamb}
\and
{ASTRON, The Netherlands Institute for Radio Astronomy, Postbus 2, 7990 AA, Dwingeloo, The Netherlands}\label{inst:astron}
\and
{Center for Astrophysics $\mid$ Harvard \& Smithsonian, 60 Garden Street, Cambridge, MA 02138, USA}\label{inst:cfa}
\and
{SRON Space Research Organisation Netherlands, Niels Bohrweg 4, 2333 CA Leiden, The Netherlands}\label{inst:sron}
\and
{University of Cambridge, Cavendish Asterophysics group, JJ Thomson Avenue, Cambridge, CB3 0HE, UK}\label{inst:cambridge}
}
 
\date{Received 18/11/2025}%; Accepted DD MMMM YYYY}

\abstract
%{Diffuse radio emission in distant galaxy clusters is preferentially revealed at low frequencies ($\nu\sim100$ MHz) due to the strong loss of energy associated with synchrotron and inverse Compton radiation that first affects the high energy particles. However, observations at these low frequencies suffer from limitations due to resolution, which affects, especially at high redshift, our ability to find the exact location of the particle acceleration and the correct separation from the compact emission from radio galaxies.}
{Enhanced inverse Compton losses at high redshift are expected to steepen the spectrum of diffuse radio sources in galaxy clusters, making low-frequency ($\nu\sim100$ MHz) observations favourable. However, observations at these low frequencies suffer from limitations due to resolution, which affects our ability to determine the exact location of the particle acceleration and the correct separation from the compact emission from radio galaxies.}
% aims heading (mandatory) we aim to unveil the properties of the radio relic in the distant galaxy cluster PSZ2\,G091.83+26.11 ($z=0.822$) by resolving the location of the particle acceleration site and  the downstream region.
{In this paper, we aim to unveil the properties of the radio relic in the distant galaxy cluster PSZ2\,G091.83+26.11 ($z=0.822$) by resolving the location of the particle acceleration site and by carefully inspect the downstream region.}
%{In this paper, we aim to identify the particle acceleration mechanism responsible for the diffuse radio emission of the radio relic in the distant galaxy cluster PSZ2\,G091.83+26.11 ($z=0.822$) by resolving the location of the particle acceleration site.}  %we aim to determine the location of the particle acceleration responsible for the emission of the radio relic in the distant galaxy cluster PSZ2\,G091.83+26.11 ($z=0.822$) and to define the kind of particle acceleration mechanism responsible for this piece of diffuse radio emission.
{We make use of the full European LOw Frequency Array (LOFAR) at 145 MHz, which enables us to study a radio relic at (sub-)arcsecond resolutions for the first time at frequencies below 1 GHz. We complement our analysis with the corresponding arcsecond-resolution observations at higher frequencies, taken with the Karl-Jansky Very Large Array (VLA).}
{We confirm that the diffuse radio emission is not associated with a radio galaxy, and the same spectral index gradient towards the cluster centre is found as in the previous $5''$-resolution maps. The $0.4''$- and $1.9''$-resolution images also reveal hints of emission ahead of the shock, with a stronger brightness in a bridge of emission connecting the relic and a radio galaxy. The $1.9''$ profiles across the relic's downstream at both 145 MHz and 3.0 GHz are well described with a log-normal distribution of magnetic fields. %($B_0=1~\mu$Gauss and $\sigma\log(B)=1$). 
The shock surface at 145 MHz presents a sharp discontinuity, in correspondence of a change in electron density, Rotation Measure and fractional polarisation values. This is possibly related with a change in the magnetic fields. Finally, we find hints of redshift evolution of the radio power versus cluster mass correlation. %that the radio power of the relic lies within the scatter of the $P_{\rm 150MHz}-E_zM_{500}$ correlation.
}
{The impressive angular resolution achievable by the LOFAR long baselines is opening an unprecedented view of the low energetic plasma in galaxy clusters. This is extremely significant in the case of high-redshift clusters, where radio emission at low frequencies is less affected by energy losses but its detection is strongly limited by poor resolution.}


\titlerunning{Sub-arcsecond LOFAR observations of PSZ2\,G091.83+26.11}
\authorrunning{G. Di Gennaro et al.}
\maketitle

%
%-------------------------------------------------------------------

\section{Introduction}
Radio relics are sources of diffuse radio emission associated with shock waves generating during merger events between two or more galaxy clusters \citep{vanweeren+19}. As merger events are the most energetic events since the Big Bang, merger shocks play a crucial role in shaping the evolution of the galaxy cluster and, at the same time, affect the microphysics of particles and magnetic field properties of the intra-cluster medium (ICM). In this context, radio relics represent the best observational evidence of the non-thermal plasma component in galaxy clusters.
They are usually co-located with X-ray shock discontinuities \citep{markevitch+vikhlinin07,botteon+18} at the cluster outskirts and are characterised by a spectral index\footnote{We define the spectral index $\alpha$ as the slope of the radio spectrum $S_\nu$, namely $S_\nu\propto\nu^\alpha$.} gradient in the radio band, from ``flatter'' ($\alpha\sim-0.8$) to ``steeper'' ($\alpha<-1.5$) values towards the cluster centre. This observational radio feature is typically associated with a Fermi-I acceleration mechanism, where particles in the ICM reach ultra-relativistic energies from scattering upstream and downstream of the shock due to magnetic inhomogeneities \citep{brunetti+jones14}. Despite the low efficiency of this mechanism in the low-Mach number regime \citep{botteon+20} typical of cluster-scale shocks \citep[i.e. $\mathcal{M}\sim1-3$;][]{markevitch+vikhlinin07}, this so-called diffusive shock acceleration \citep[DSA;][]{blandford+eichler87} mechanism is usually invoked to explain the presence of radio relics. To overcome the unrealistically high efficiencies that are often required to explain the high radio luminosity when assuming the DSA of thermal electrons, the presence of fossil plasma (such as those ejected in the ICM by cluster radio galaxies or phoenices) is usually invoked \citep{markevitch+05,kang+12,kang+17}. Although the detection of connections between radio galaxies and diffuse radio emission is not strictly required, only a handful of clear cases of re-acceleration have been observed so far \citep{bonafede+14,vanweeren+17a,digennaro+18}.

The LOw Frequency ARray \citep[LOFAR,][]{vanhaarlem+13} has been groundbreaking in the study of diffuse radio emission in clusters, combining high sensitivities ($\rms$) and resolutions ($\Theta$) in the low (120--168 MHz) and ultra-low (20--68 MHz) frequency regimes.
In particular, the LOFAR Two-Meter Sky Survey \citep[$\rms=100~\mu$Jy\,beam$^{-1}$ and $\Theta=6''$ in the 120--168 MHz frequency band;][]{shimwell+17,shimwell+19,shimwell+22} has observed a plethora of new radio sources, especially identifying those with ultra-steep spectra (i.e. $\alpha<-1.5$). These sources are associated with intrinsically low-energy processes or have aged and lost their energy through synchrotron and inverse Compton processes \citep{cassano+brunetti05}. Due to their steep spectrum and low surface brightness, these sources are usually undetected at higher frequencies (i.e. $\nu>1$ GHz). 

\begin{table}
\caption{Cluster information.}
\vspace{-5mm}
\begin{center}
\resizebox{0.5\textwidth}{!}{
\begin{tabular}{lr}
\hline
\hline
Cluster name & PSZ2\,G091.83+26.11 \\
Redshift ($z$) & 0.822 \\
Right Ascension (RA) & $\rm 18^h31^m11.136^s$\\
Declination (DEC) & $+62^\circ14^\prime56.04''$\\
SZ Intensity ($Y_{\rm SZ,5R500}~[\times10^{-3}~{\rm arcmin^2}]$) & $1.0\pm0.1$ \\
Mass ($M_{\rm SZ,500}~[\times10^{14}~{\rm M_\odot}$]) & $7.4\pm0.4$ \\
Relic flux density at 144 MHz ($S_{\rm RELIC,144}$ [mJy]) & $248.8\pm40.9$ \\
Relic flux density at 1.5 GHz ($S_{\rm RELIC,1518}$ [mJy]) & $15.9\pm0.6$ \\
Relic flux density at 3.0 GHz ($S_{\rm RELIC,2997}$ [mJy]) & $5.0\pm0.2$ \\
Cosmological scale ($\rm kpc/''$) & 7.576 \\
\hline
\end{tabular}
}
\end{center}
\vspace{-5mm}
\tablefoot{Cluster mass and SZ intensity are taken from \cite{planckcoll16}. The relic flux densities at 144 MHz and 3.0 GHz are those reported in \cite{digennaro+23}, at $12''$ resolution.}
\label{tab:cluster}
\end{table}

\begin{table*}[h!]
\caption{International LOFAR Telescope observation details.}
\vspace{-5mm}
\begin{center}
\resizebox{0.7\textwidth}{!}{
\begin{tabular}{lccccc}
\hline
\hline
Project & Field & Separation & Observation date & Observation length & No. stations \\
& & [deg] & [dd/mm/yyyy] & [h] & core/remote/international\\ 
\hline
LC06\_015 & P275+63 & 1.1 & 22/08/2016 & 8 & 48/14/12 \\
LC20\_022 & N/A & 0 & 23/11/2024 & 8 & 46/13/14 \\
\hline
\end{tabular}
}
\end{center}
\vspace{-5mm}
\tablefoot{The observations during the project code LC20\_022 were centred on the target. The observation length refers to the on-target time.}
\label{tab:obs}
\end{table*}

Among the discoveries, LoTSS observations have revealed the presence of extended diffuse radio emission in distant clusters, when the Universe was only about half of its current age \citep{cassano+19,digennaro+21a,digennaro+25a}. As for diffuse radio sources in the local Universe, they are  associated with merger-induced turbulence and shock acceleration \citep{digennaro+21c}. However, at these redshifts (i.e. $z>0.6$) the limitation in resolution has challenged the proper subtraction of compact emission associated with cluster galaxies \citep{osinga+20,digennaro+25a}, as well as the proper detection of these sources. The latter challenge is particularly important for radio relics, as their detection could be missed due to the effect of poor angular resolution and projection effects \citep{nuza+17}.
In this context, the International LOFAR Telescope (ILT), which currently includes baselines up to 2,000 km, represents a very effective instrument to combine high-resolution and high sensitivity \citep[about $\rm35~\mu Jy\,beam^{-1}$ for a standard 8-hour observation, see][]{morabito+22,sweijen+22} below $200$ MHz frequencies. The nominal resolution of the ILT 120--168 MHz observations is $0.3''$, which has already shown the capability to disentangle compact from extended radio emission in distant clusters \citep{hlavacek-larrondo+25}, as well as filaments in diffuse radio emission in clusters \citep{vanweeren+24} and in cluster radio galaxies \citep{timmerman+22,timmerman+24,cordun+23,pasini+25,derubeis+25,morabito+25}. 

In this work, we present the observations of PSZ2\,G091.83+26.11 (hereafter \cluster) at $z=0.822$ (Tab.~\ref{tab:cluster}). The extended radio emission in the cluster, namely a central radio halo\footnote{Radio halos are an additional class of diffuse radio emission that are associated with cluster mergers \citep{brunetti+jones14}. We refer to \cite{brunetti+jones14} and \cite{vanweeren+19}, and references therein, for theoretical and observational reviews on all the kinds of diffuse radio sources in clusters.} and an eastern radio relic, were previously reported in \cite{digennaro+21a,digennaro+21c,digennaro+23} using the LOFAR high-band antennas (HBA, 120--168 MHz), the upgraded Giant Metrewave Radio Telescope \citep[uGMRT;][]{gupta+17} in Band 3 (250--500 MHz) and Band 4 (550--900 MHz), and the Karl Jansky Very Large Array \citep[VLA;][]{perley+11} in L- (1--2 GHz) and S-band (2--4 GHz).
Despite the large bandwidth coverage, a detailed analysis of the radio relic in the cluster is still missing due to the limitation in resolution at low frequencies (i.e. $\nu<1$ GHz). Previous spectral index analyses were limited by the angular resolution of the LOFAR 120--168 MHz observations (i.e. $\Theta\sim5''-6''$) which, at the cluster redshift, corresponds to a physical scale of $\sim40$ kpc. 

The paper is structured as follows: in Sect. \ref{sec:observations} we highlight the long-baseline observations and data reduction; in Sect. \ref{sec:results} we show the novel high-resolution images and spectral index maps; in Sect. \ref{sec:discussion} we discuss the new features and the implications in the context of previous low-redshift studies; we summarise our results and conclusions in Sect. \ref{sec:conclusion}.

\begin{figure}
\centering
\includegraphics[height=0.24\textwidth]{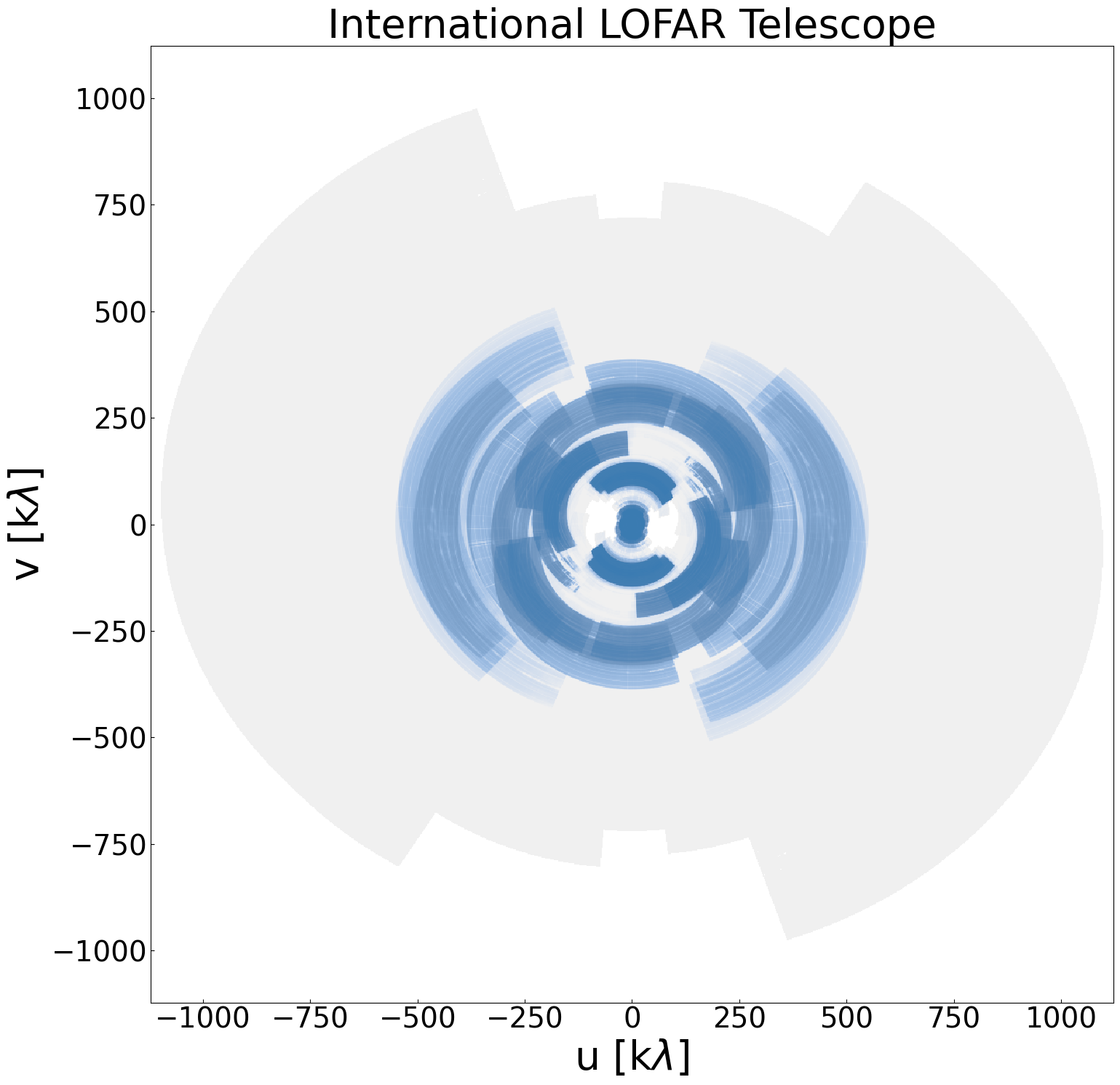}%{ilt.png} 
\includegraphics[height=0.24\textwidth]{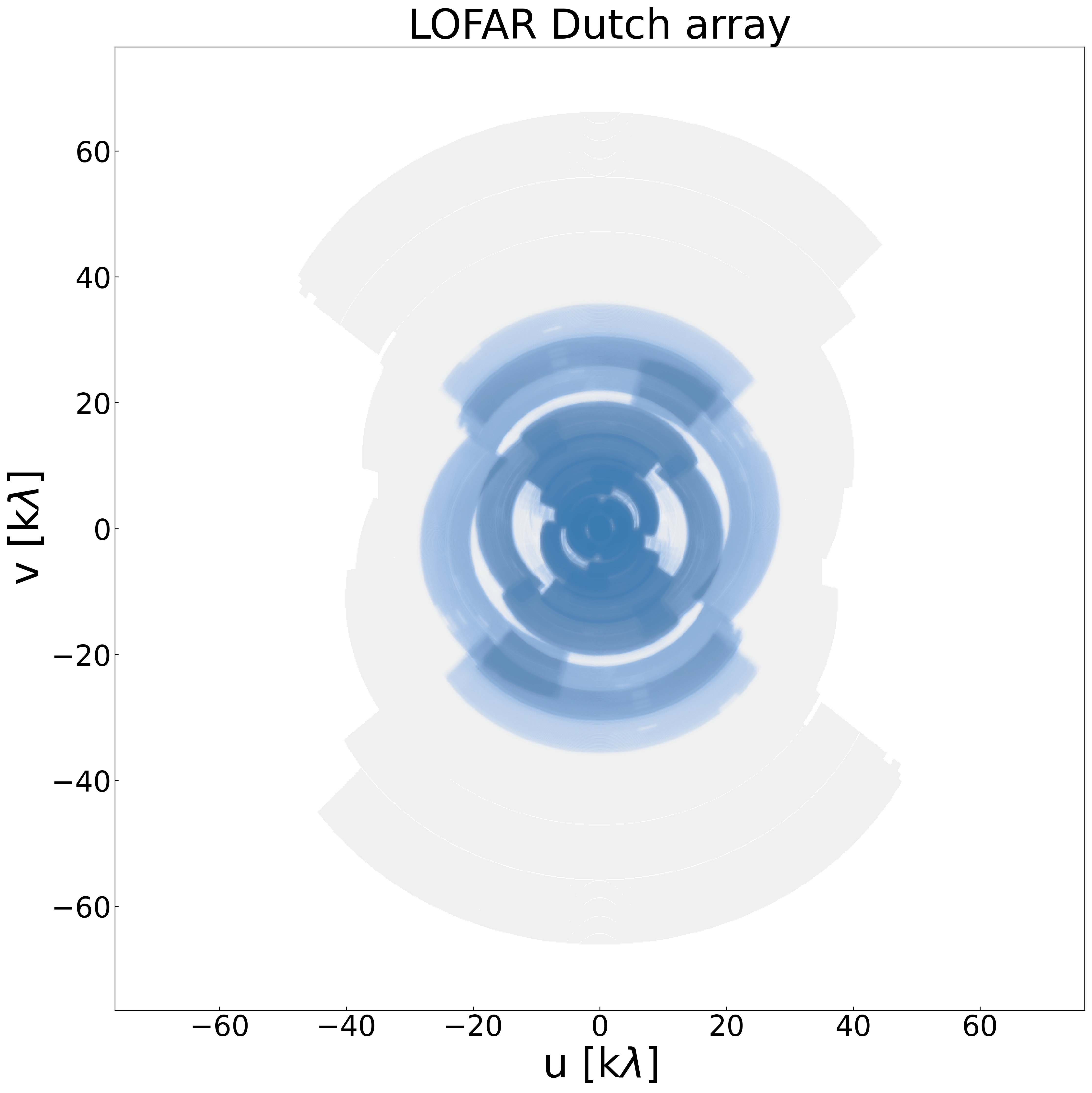}%{hba.png} 
\vspace{-3mm}
\caption{Comparison of the {\it uv}-coverage for the LOFAR 145 MHz long-baseline (left panel) and the LOFAR 145 MHz Dutch array (right panel).}\label{fig:uvcoverage}
\end{figure}

\section{Observations and data reduction}\label{sec:observations}
In this section we present the observations and data reduction of the novel ILT observations and the present the ancillary data used in the analysis.

\subsection{International LOFAR Telescope (ILT)}
\cluster\ was observed as part of the LOFAR Two-Meter Sky Survey \citep[LoTSS;][]{shimwell+17,shimwell+19,shimwell+22}, which simultaneously collects data from Dutch and international stations. The cluster area is covered by four LoTSS pointings, namely P275+60, P281+63, P275+63, and P280+60. Among these, only P275+60 could be used for ILT analysis, with \cluster\ located inside the full width half maximum (FWHM) of international station beam \citep[i.e.  $\rm FWHM_{ILT}\sim1.25^\circ$, instead of the $\rm FWHM_{LoTSS}=2.0^\circ$ of the Dutch stations, and the cluster at a distance of $1.1^\circ$ from the pointing centre, see][]{morabito+22,dejong+24}.
To improve the detection of cluster radio emission, which was highly affected by time and bandwidth smearing, we obtained additional targeted observations (project code: LC20\_022; PI: G. Di Gennaro). The final dataset includes these on-target observations as well as the data from the pointing P275+60 (see Tab. \ref{tab:obs}), for a total of 16 hours on target. The comparison of the {\it uv}-coverage of the LOFAR Dutch stations and the LOFAR long baseline is shown in Fig.~\ref{fig:uvcoverage} (right and left panel, respectively).

The first step for the calibration strategy is to process the baselines associated with the Dutch LOFAR observations stations. 
These data were calibrated using the {\tt LINC} pipeline\footnote{\url{https://git.astron.nl/RD/LINC}} \citep{williams+16,vanweeren+16,degasperin+19}. This pipeline first takes data from the 10-minute scan on the primary calibrator observed directly before or after both target observations and uses these to derive calibration solutions for the polarization alignment, bandpass and clock offsets for each station per observation. Next, these calibration solutions were applied to the target data, after which an initial direction-independent phase calibration was determined only for the Dutch stations using a sky model obtained from the TIFR Giant Metrewave Radio Telescope Sky Survey \citep[TGSS,][]{intema+17}. This phase calibration was then refined using the {\tt DDF} pipeline\footnote{\url{https://github.com/mhardcastle/ddf-pipeline}} \citep{tasse+21}, which uses {\tt DDFacet}\footnote{\url{https://github.com/saopicc/DDFacet}} \citep{tasse+18} and {\tt killMS}\footnote{\url{https://github.com/saopicc/killMS}} \citep{tasse14,smirnov+tasse15} to determine direction-dependent phase calibration solutions for the Dutch stations to account for the varying ionosphere across the field of view.

\begin{table}%[h!]
\caption{Radio imaging details.}
\vspace{-5mm}
\begin{center}
\resizebox{0.45\textwidth}{!}{
\begin{tabular}{ccccc}
\hline
\hline
Central frequency & Resolution & {\it uv}-taper & Robust & Map noise \\
$\nu$ [MHz] & $\Theta$ [$''\times'',^\circ$] & [$''$] & & $\rms$ [$\mu$\jybeam] \\
\hline
145  & $0.4\times0.3$, 163 & - & $-0.5$ & 25.4 \\
     & $1.8\times1.0$, 110 & 1.2 & $-1.5$ & 91.6 \\
     & $3.0\times1.7$, 111 & 1.5 & $-1.5$ & 120.5 \\
     & $3.8\times2.7$, 103 & 2.0 & $-1.5$ & 148.2 \\
     & $5.0\times4.4$, 94 & 4.0 & $-1.5$ & 134.0 \\
%650  & $4.1\times2.4$, XX & - & $-1.25$ & XX \\
1518 & $2.9\times2.2$, 60 & - & $-1.25$ & 13.6 \\
%     & $4.3\times3.2$, 67 & - & 0 & 8.3 \\
2997 & $1.8\times1.1$, 60 & - & $-1.25$ & 3.8 \\
     & $2.9\times2.2$, 75 & - & 0 & 2.4 \\
\hline
\end{tabular}
}
\end{center}
\vspace{-5mm}
\tablefoot{The observations at 1.5 and 3.0 GHz (VLA) were only used for the spectral index analysis, after convolving to the same beam size and aligning to the same pixel scales \citep{digennaro+23}.}
\label{tab:images}
\end{table}

To extend the calibration to the international LOFAR stations and reach angular resolutions better than $6''$, we applied the {\tt LOFAR-VLBI} pipeline\footnote{\url{https://github.com/lmorabit/lofar-vlbi}; this was used at the time of the calibration of this work; a recent upgrade of the Pipeline for the International LOFAR Telescope ({\tt PILoT}) described in van der Wild et al., submitted.} \citep{morabito+22} to the data. First, we selected a bright and compact radio source within the field of view  (WISE J183124.46+623034.4), that is the in-field calibrator, as provided by the Long Baseline Calibrator Survey \citep[LBCS;][]{jackson+22}. Then, we identified a region containing both the in-field calibrator source and the target, and we subtracted all the $6''$-LoTSS emission outside of this region from the data. After that, we derived calibration solutions for the international stations using the in-field calibrator by means of {\tt facetselfcal}\footnote{\url{//github.com/rvweeren/lofar_facet_selfcal}} \citep{vanweeren+21}.

With all calibration solutions applied to the data, the final ILT images were produced using {\tt WSclean v3.4} \citep{offringa+14,offringa+17} using {\tt Briggs} weighting and {\tt robust -0.5} for the highest resolution images. We also produced lower-resolution images, using {\tt robust -1.5} and different tapers (i.e. {\tt taper=1.2$''$, 1.5$''$, 2.0$''$, 4.0$''$}, see Tab.~\ref{tab:images}). All the ILT images presented in this work have a central frequency of 145 MHz. The highest-resolution image has a beam size of $\Theta=0.4''\times0.3''$ and a noise level of $\rms=25.4~\mu$\jybeam. 

\subsection{Ancillary high-frequency data}
We made use of Karl Jansky Very Large Array (VLA) observations in L- (1--2 GHz) and S-band (2--4 GHz) in B+C+D configuration (Tab.~\ref{tab:images}), for a total time of 15 hours for each band. These were already published in \cite{digennaro+23}, therefore we point to that work for the description of the data reduction.

\begin{figure}
\centering
\includegraphics[width=0.5\textwidth]{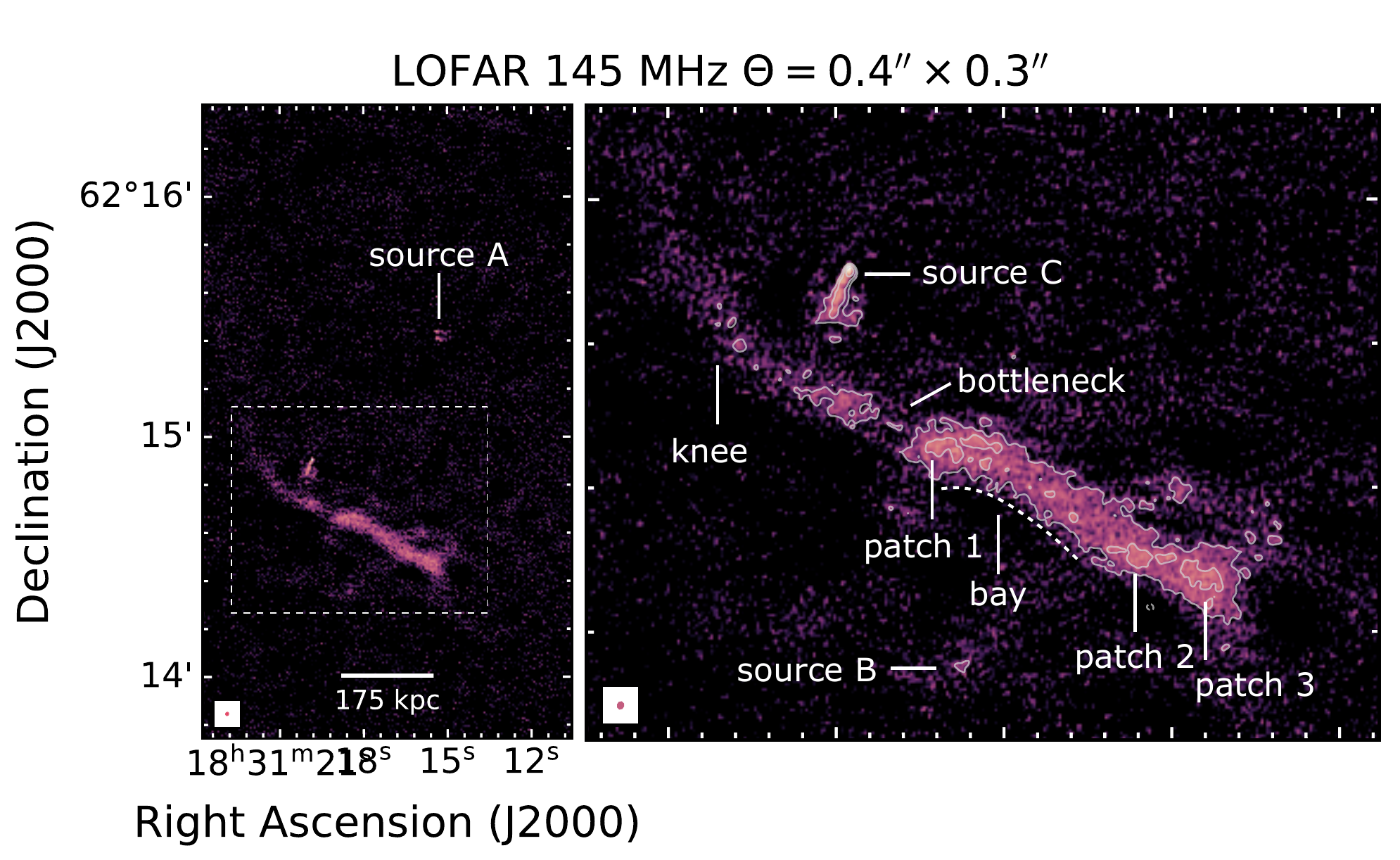} 
\vspace{-7mm}
\caption{Full-resolution (i.e. $\Theta=0.4''\times0.3''$) ILT 145 MHz images of \cluster\ centred on the radio relic. On the left panel, we show the whole radio relic area, while in the right panel we show a zoom on R2, with radio contours drawn starting from $2\rms$ (where $\rms=25.4$ $\mu$\jybeam) and the new features highlighted.}\label{fig:ilt-images-high}
\end{figure}

\begin{figure*}
\centering
\includegraphics[width=\textwidth]{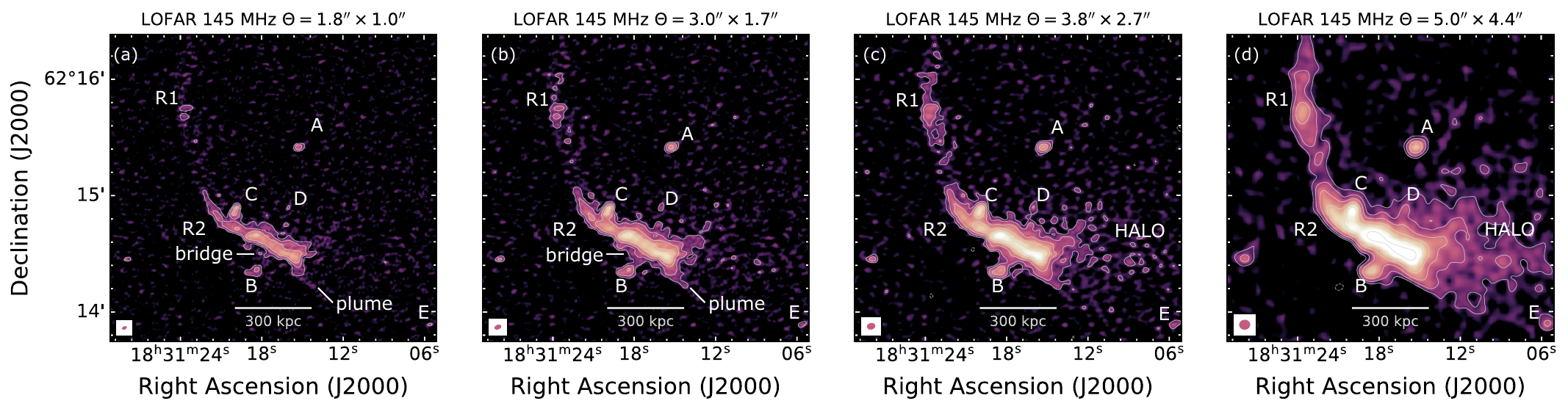} 
\vspace{-7mm}
\caption{Lower-resolution ILT 145 MHz images of \cluster. From left to right: $\Theta=1.8''\times1.1''$ ($\rms=91.2$ $\mu$\jybeam); $\Theta\sim3.0''\times1.7''$ ($\rms=120.5$ $\mu$\jybeam); $\Theta\sim3.6''\times2.9''$ ($\rms=148.2$ $\mu$\jybeam);  $\Theta\sim5.0''\times4.3''$ ($\rms=134.0$ $\mu$\jybeam). Radio contours are drawn at $2.5\rms\times[-1,1,2,4,8,16,32]$, with $\rms$ the noise level in each image (negative contours are drawn with dashed line). Source labelling follow \cite{digennaro+21c}.}\label{fig:ilt-images-low}
\end{figure*}

\section{Results}\label{sec:results}
In Figures \ref{fig:ilt-images-high} and \ref{fig:ilt-images-low}, we show the images of \cluster\ made with the LOFAR long baselines at different resolutions (i.e. $0.4''\times0.3''$, hereafter high resolution, Fig.~\ref{fig:ilt-images-high}; $1.8''\times1.1''$, $3.0''\times1.7''$, $3.8''\times2.7''$, and $5.0''\times4.4''$, Fig.~\ref{fig:ilt-images-low}a--d). At sub-arcsecond resolutions, only the southern part of the radio relic (R2) and the radio galaxies labelled as A, B and C are detected (see Fig. \ref{fig:ilt-images-high}). As opposed to what was previously observed at higher frequencies and lower resolution \citep{digennaro+21c,digennaro+23}, the structure of R2 is not sharp, but a bay of emission is present in the northern part (see dashed line in Fig. \ref{fig:ilt-images-high}, right panel). The radio emission is also not completely uniform: from south-east to north-west, we see three patches of emission at the $4\rms$ noise level, a discontinuity (``bottleneck'', which becomes less prominent at $1.8''\times1.0''$ resolution) and a deviation towards the north (``knee''; see the right panel in Fig.~\ref{fig:ilt-images-high}). The sub-arcsecond resolution image also reveals a more complex structure for the radio galaxies than previously observed, with sources A and C having a double-lobe and a head-tail morphology, respectively.

\begin{table}%[h!]
\caption{Flux densities at $3.0''$ resolution.}
\vspace{-5mm}
\begin{center}
\resizebox{0.45\textwidth}{!}{
\begin{tabular}{ccccc}
\hline
\hline
Source name & Frequency & Flux density & Spectral index \\
& $\nu$ [MHz] & $S_\nu$ [mJy] & $\alpha$ \\
\hline
source A & 145 & $3.8\pm0.5$ & $-0.67\pm0.06$ & \\
& 1518 & $1.2\pm0.1$ \\
& 2997 & $0.53\pm0.03$ \\
source B & 145 & $6.4\pm0.8$ & $<-2.2$ & \\
& 1518 & $<0.05$ \\
& 2997 & $<0.008$ \\
source C & 145 & $12.6\pm1.3$ & $-0.75\pm0.03$ & \\
& 1518 & $4.7\pm0.2$ \\
& 2997 & $1.9\pm0.1$\\
R1 & 145 & $15.1\pm2.1$ & $-1.07\pm0.05$ & \\
& 1518 & $1.0\pm0.2$ & \\
& 2997 & $0.60\pm0.04$ & \\
R2 & 145 & $168.4\pm16.9$ & $-1.25\pm0.04$ & \\
& 1518 & $11.4\pm0.6$ \\
& 2997 & $3.8\pm0.2$ \\
\hline
\end{tabular}
}
\end{center}
\label{tab:fluxes}
\end{table}

The additional known radio features appear only at lower resolutions ($\Theta\gtrsim2.0''$; see Fig.~\ref{fig:ilt-images-low}a--d). At these resolutions, source B appears to be more diffuse than previously observed. This source was also detected at 650 MHz by uGMRT observations \citep{digennaro+21c} but not in higher-frequency VLA observations \citep{digennaro+23}, indicating a steep spectrum \citep[$\alpha<-2$;][]{digennaro+21c}. We additionally detect faint diffuse emission at LOFAR frequencies that connects source B with the brightest part of the radio relic (i.e. R2). This bridge of emission could provide a possible contribution of plasma to the relic and could justify the difference in brightness of R2 compared to R1. At lower resolutions, this emission covers entirely the upstream region of R2, in a $\rm 30\times300~kpc^2$ region ahead of the relic.
At $\Theta=3.8''\times2.7''$, also the faintest part of the relic (R1) is detected, as well as an extension of R2 south-eastward (``plume'') which then blends with the radio halo. This Mpc-scale, approximately round emission appears clearly at lower resolutions \citep[i.e. $\Theta=5.0''\times4.3''$, comparable with the highest resolution reached with the Dutch stations only, see][]{digennaro+21a}.

\begin{figure}
\centering
\includegraphics[width=0.45\textwidth]{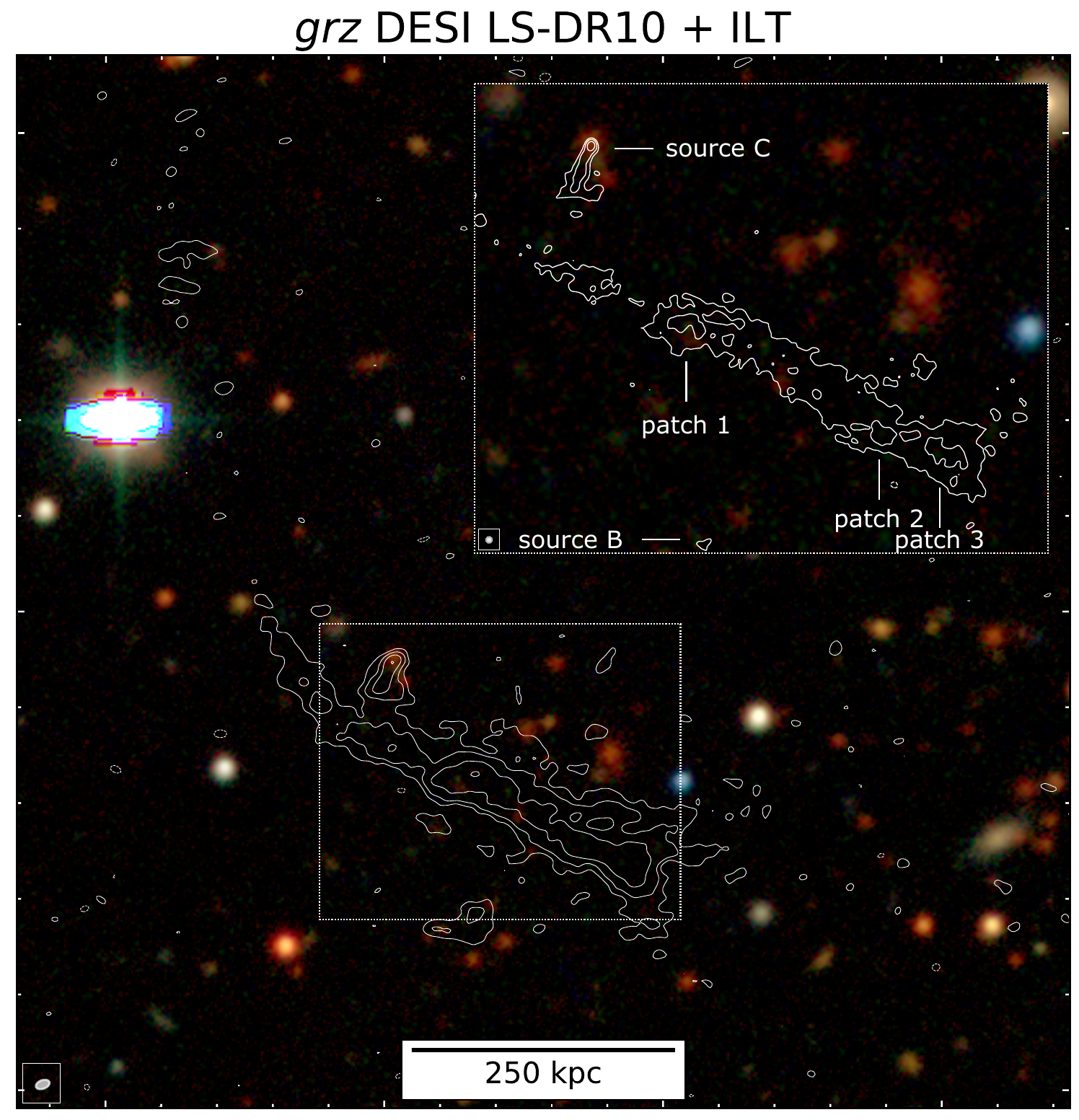}
\vspace{-3mm}
\caption{{\it grz} image from the 10th data release of DESI Legacy Survey (DR10) of \cluster\ with the $1.8''$ ILT radio contours. In the inset, we show the zoom-in on R2, with radio contours at $0.4''$ resolution. For both radio images, white contours are at the $2.5\rms\times[1,2,4,8,16]$ levels (see Tab.~\ref{tab:images} for the map noise at these resolutions), and the beam size is displayed in the lower left corners.}
\label{fig:opt-radio}
\end{figure}

\begin{figure*}
\centering
\includegraphics[height=6cm]{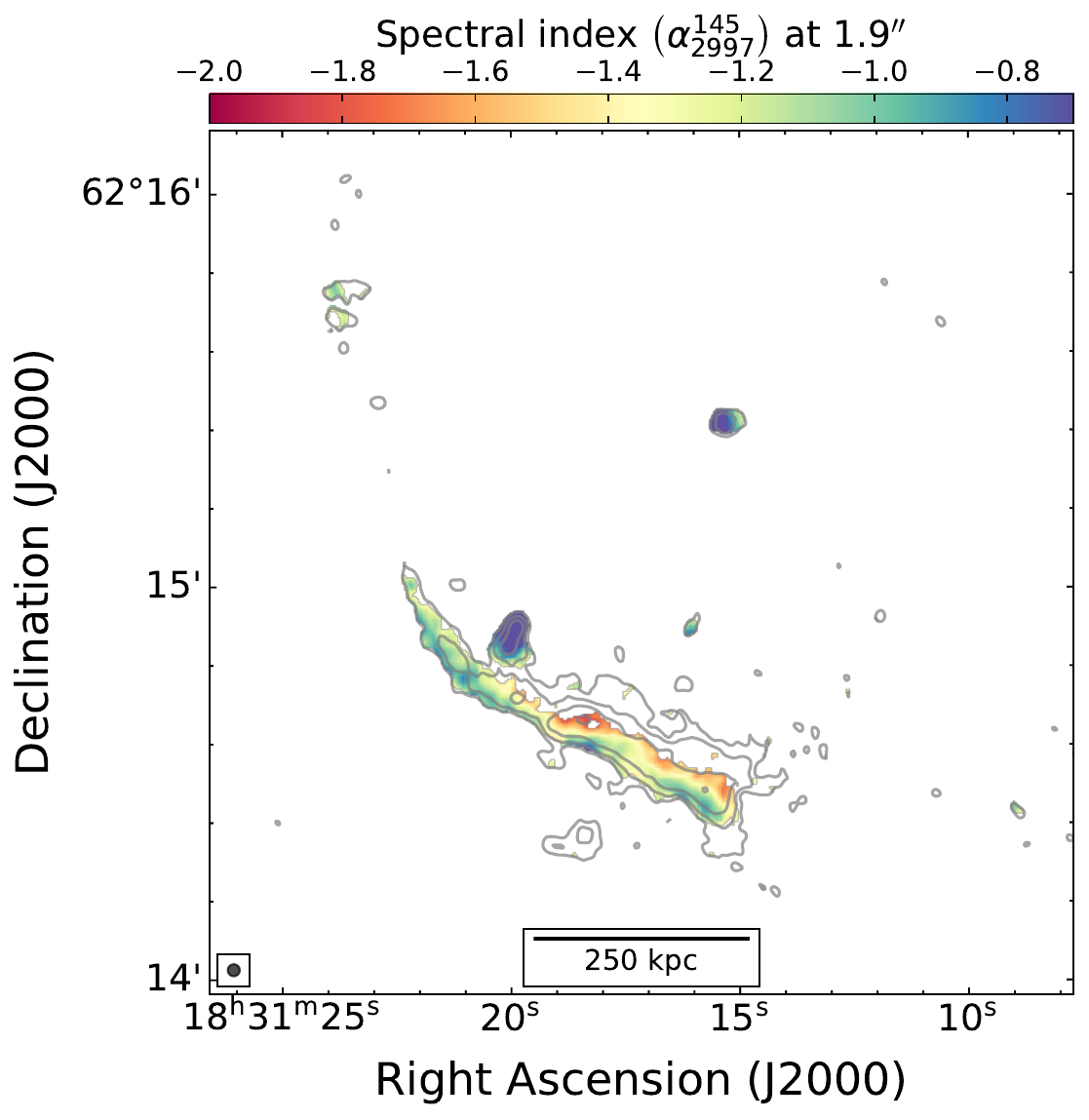}
\includegraphics[height=6cm]{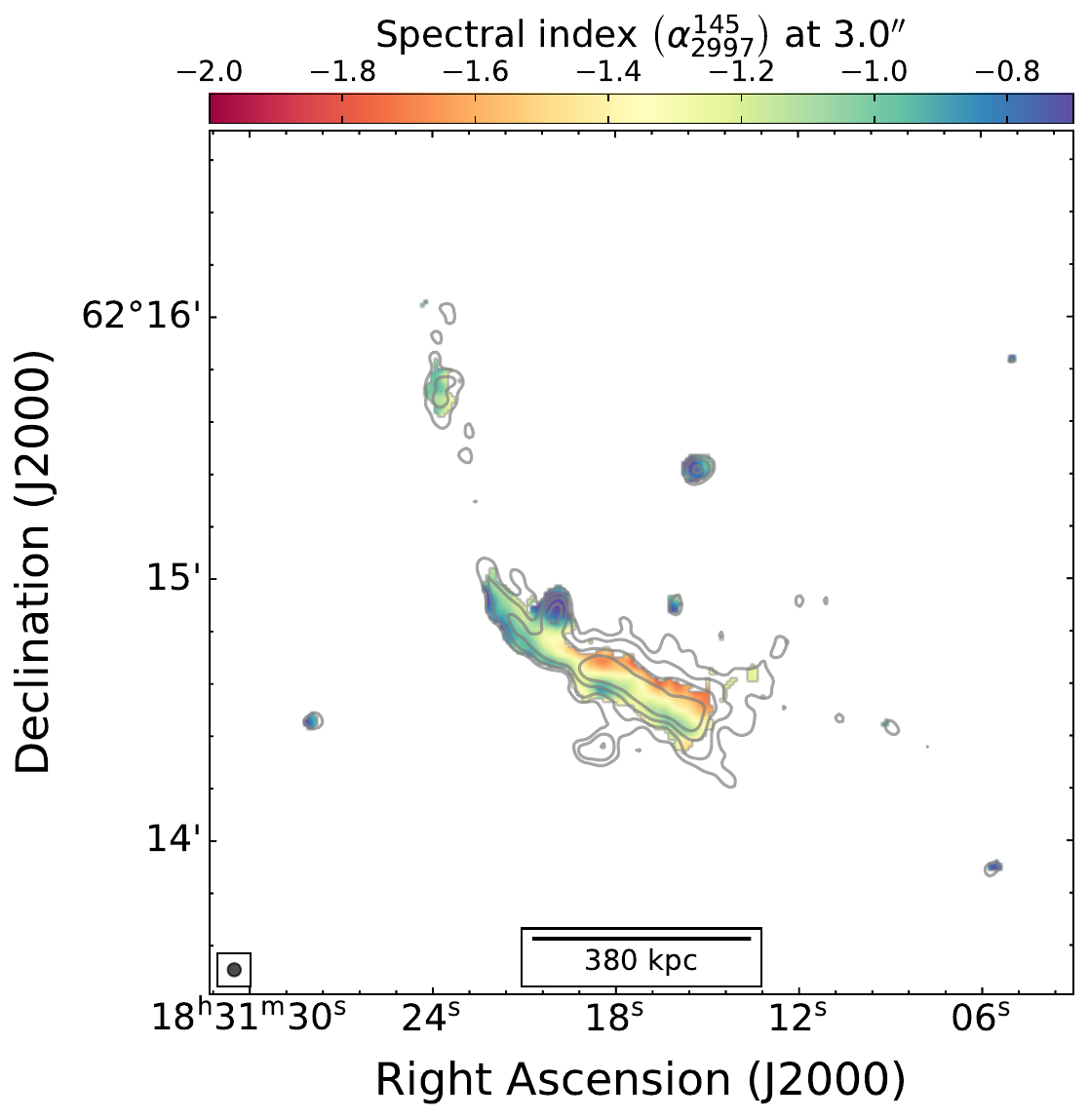}
\includegraphics[height=6cm]{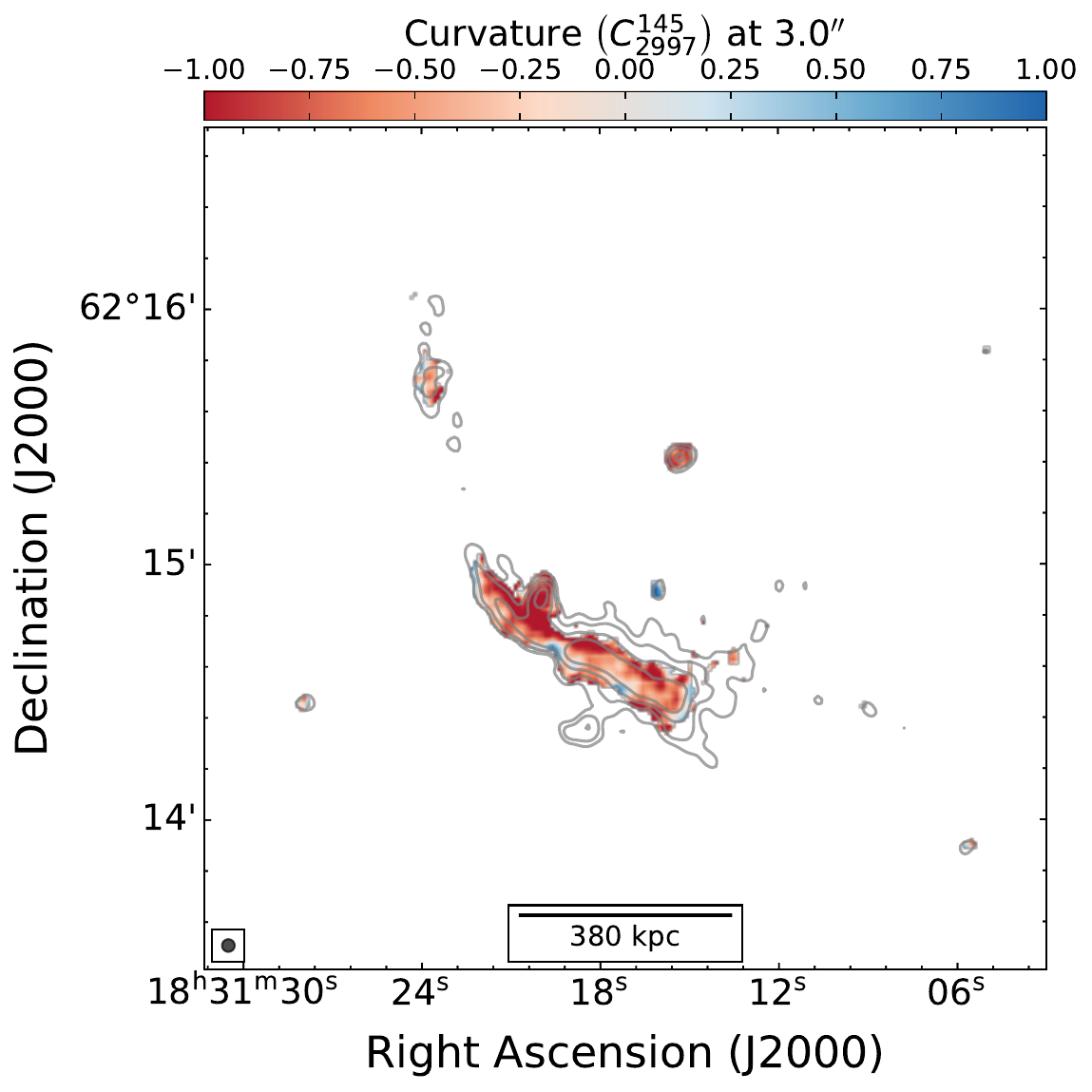}
\vspace{-3mm}
\caption{Spectral analysis of \cluster. Left: spectral index map at $1.9''$, using 145 MHz and 3.0 GHz observations. Central: spectral index map and $3.0''$, using 145 MHz, 1.5 GHz and 3.0 GHz observations. Right: spectral curvature map at $3.0''$. Radio contours in each panel are taken from the ILT 145 MHz at the corresponding resolution, and are drawn starting from the $3\rms$ level.}\label{fig:spixmpas}
\end{figure*}

The lack of point-like structures at high and intermediate resolutions that can be associated with optical galaxies in Fig.~\ref{fig:opt-radio} disfavours the possibility that this elongated diffuse emission is associated with a radio galaxy and supports the scenario of extended diffuse radio emission proposed with high-frequency observations \citep{digennaro+23}. Interestingly, we note that the patches at the $4\rms$ level in R2 also do not have a clear optical counterpart, meaning that more likely they are tracing substructures in the relic.
Similarly, the central diffuse radio emission is undetected at high resolution and does not show any coincidence with the optical galaxies, strongly suggesting that its origin is not related to unresolved active galactic nuclei (AGN) emission, but it is indeed diffuse radio emission. In Tab.~\ref{tab:fluxes}, we reported the flux densities at $3.0''$ resolution. This resolution was chosen to detect and separate the contribution of all the relevant compact radio sources, that are source A, B and C, and the two relic components, that are R1 and R2.  We also reported the integrated radio spectral index (see Sect. \ref{sec:spix} and Eq. \ref{eq:fitalpha})

\subsection{Spectral and curvature index maps}\label{sec:spix}
Given these new LOFAR high-resolution images, we can produce for the first time a spectral index analysis of a cluster at high redshift at the arcsecond-scale, i.e. tracing particles on the physical scale of $\sim15$ kpc, across a wideband frequency (i.e. from 145 MHz up to 3.0 GHz).

We created a spectral index map of the cluster at both $1.9''$ and $3.0''$ resolutions. 
In the former case, only the ILT 145 MHz and VLA 3.0 GHz data could be used, while for the latter we also added data from the 1.5 GHz VLA observations (see Tab.~\ref{tab:obs}). All images were convolved to the same beam size ($\Theta$) and aligned to the same pixel scale ($d\Theta$) and astrometry \citep[as described in][]{degasperin+23}. The final maps are shown in Fig.~\ref{fig:spixmpas}, and are obtained by fitting a first-order polynomial in logarithmic space, that is 
\begin{equation}\label{eq:fitalpha}
\log S_\nu = \log S_0 + \alpha^{145}_{2997}\log\nu \, ,
\end{equation}
where $\alpha^{145}_{2997}$ is the spectral index calculated between the lowest and highest available frequency, and $S_0$ the normalisation, for each pixel above the $3\rms$ level. We also present the curvature map at $3.0''$ resolution (Fig.~\ref{fig:spixmpas}, right panel), obtained as a comparison of the spectral indices calculated between 145 MHz and 1.5 GHz, and between 1.5 GHz and 3.0 GHz (i.e. $C=\alpha^{1518}_{2997} - \alpha^{145}_{1518}$). In this case, we calculate the spectral index using the analytical expression, that is 
\begin{equation}\label{eq:analyticalpha}
\alpha^{\nu_1}_{\nu_2}=\log_{10} \left ( \frac{S_{\nu_1}}{S_{\nu_2}}\right ) / \log_{10} \left ( \frac{\nu_1}{\nu_2}\right ) \, ,
\end{equation}
with $\nu_1$ and $\nu_2$ the lower and higher frequency respectively and $S_{\nu_1}$ and $S_{\nu_2}$ their correspondent surface brightness values. Under this assumption, negative curvature values correspond to a spectrum with a steeper spectral index at high frequencies.

We find a similar spectral behaviour at both $1.9''$ and $3.0''$ resolutions, with the radio galaxies having a flat spectral index (i.e. $\alpha^{145}_{2997}\sim-0.7$) and the brightest part of the radio relic having a spectral index gradient towards the cluster centre, from $\alpha^{145}_{2997}\sim-1$ to $\alpha^{145}_{2997}\sim-2$. This gradient is in agreement with what was found at lower resolutions \citep[$\Theta>5''$;][]{digennaro+23} and is consistent with the spectral index uncertainties, which are obtained by 150 Monte Carlo realisations of the fit in Eq. \ref{eq:fitalpha}. The high resolution in hand allowed us to identify spectral index variations in the outermost region of the relic, with a steepening at $\sim120$ kpc from the south-west end of the relic (i.e. $\alpha^{145}_{2997}\sim-1.3$).
This region was identified as a ``break'' with the VLA 3.0 GHz observations in \cite{digennaro+23}, and it is also the location where the bridge of radio emission from source B connects to the relic at lower resolutions. 
The relic is mainly characterised by negative curvature values ranging from $C^{145}_{2997}\sim-0.2$ to $C^{145}_{2997}<-1$. These values are significant considering the uncertainties calculated using error propagation, that is $\Delta C^{145}_{2997}=\sqrt{\left ( \Delta\alpha^{1518}_{2997} \right )^2 + \left ( \Delta\alpha^{145}_{1518} \right )^2}$, where $\Delta\alpha^{\nu_1}_{\nu_2}=\frac{1}{\ln(\nu_2/\nu_1)}\sqrt{\left ( \frac{\Delta S_{\nu_1}}{S_{\nu_1}} \right )^2 + \left ( \frac{\Delta S_{\nu_2}}{S_{\nu_2}} \right )^2}$. Notably, negative curvature values are also found for sources A and C.

\begin{figure*}
\centering
\includegraphics[width=\textwidth]{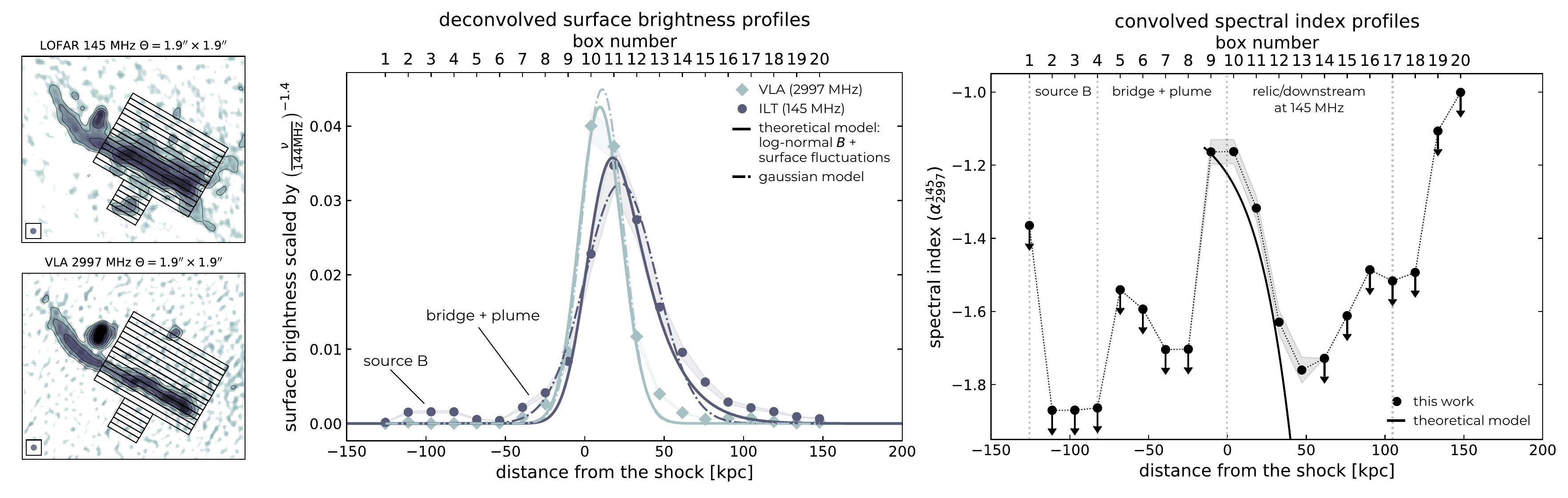}
\vspace{-7mm}
\caption{Radio profiles across the relic at $1.9''$. Left panel: sectors from the 145 MHz and 3.0 GHz images (top and bottom, respectively) used to derive the profiles. Central panel: normalised 145 MHz and 3.0 GHz deconvolved surface brightness profiles (dark circles and light diamonds, respectively); solid lines represent the theoretical model, assuming a log-normal distribution of the magnetic fields ($B_0=1~\mu$G and $\sigma=1$), while dot-dashed lines show the Gaussian profile. Right panel: convolved spectral index profiles, with the theoretical model overlaid as black line; dotted vertical lines delimit different regions of interest.}\label{fig:profiles}
\end{figure*}

\begin{comment}
\subsubsection{Spectral tomography}
We also investigated possible overlays of different filaments and substructures in the relic using the spectral tomography technique \citep{katz-stone+rudnick97b}. According to this, filaments and substructures with different spectral behaviour are easily recognised by subtracting the image at frequency $\nu_2$ scaled with a given spectral index $\alpha_i$ from the image at frequency $\nu_1$:
\begin{equation}
I(\alpha_i) = I_{\nu_1} - \left ( \frac{\nu_1}{\nu_2} \right )^{\alpha_i}\, I_{\nu_2} \, .
\end{equation}
In this way, positive and negative values describe structures with steeper and flatter spectral indices than $\alpha_i$, respectively, while structures with $\alpha=\alpha_i$ disappear from the resulting image \citep{katz-stone+rudnick97a,crawford+01,delaney+02,gizani+leahy03,mckinley+13,rajpurohit+21,brienza+25}.

In the case of \cluster, we use $\nu_1=145$ MHz and $\nu_2=3.0$ GHz at $1.9''$ resolution, and varied $\alpha_i$ from $-0.8$ to $-1.7$ in steps of 0.1 (see Fig.~\ref{fig:tomography}). From Fig.~\ref{fig:tomography}, it is clear that the entire relic in \cluster\ is overall characterised by a spectral index of $\sim -1.1$, as also found in the spectral index map (see Fig.~\ref{fig:spixmpas}). However, we also identify spots of flatter/steeper spectral indices in R2 (see black and white arrows, respectively). We also detect hints of a filament between source B and R2, as long as R2, and a plume of emission from the southwestern edge of the relic, which disappear at $\alpha_i<-1.5$.

\end{comment}

\section{Discussion}\label{sec:discussion}
The new ILT observations of \cluster\ allow us to investigate the nature of a distant radio relic in unprecedented detail. In particular, we focus on the radio emission in R2, which is resolved at sub-arcsecond resolution.

\begin{comment}
Why R2 is more bright than R1? 
- comparison of a "fit" for the profiles with a/multiple normal and with a log-normal functions \\
- fossil plasma coming from source B \\
- superimposition of a radio galaxy (FRI) -- but what about the optical counterpart? are the radio peaks diffuse hot spots?
- "pre"-shock? 
\end{comment}

\subsection{Radio profiles}
In order to investigate the properties of the radio relic in \cluster, we extracted the surface brightness profiles of the radio relic from the  $1.9''$-resolution convolved images at 145 MHz and 3.0 GHz, which additionally enable us to also perform a spectral index profile analysis (see Sect. \ref{sec:downstream}). Additional high-resolution information on the radio shock surface was obtained from the $0.4''$-resolution image at 145 MHz (see Sect. \ref{sec:shock}).

\subsubsection{Relic downstream}\label{sec:downstream}
To investigate the downstream properties of the radio relic and the possible contribution of fossil plasma to its radio power, we extracted surface brightness profiles from rectangular sectors with the shape of $1.9''\times9''$ to cover the emission of source B and $1.9''\times40''$ to cover the radio relic and the emission in the relic's upstream (i.e. the bridge and the plume). These values correspond to the beam size and to the length of source B and R2, respectively (see Fig.~\ref{fig:profiles}, left panel). To reduce the effects of beam convolution, we extracted the surface brightness profiles from the clean component model images \citep{vanweeren+10,lusetti+25}. 

The normalised deconvolved surface brightness profiles are shown in Fig.~\ref{fig:profiles} (central panel). We shifted the profiles to match a semi-analytic description, see discussion below. In this convention, the outermost edge of the relic is located in the correspondence of box \#10, negative physical distances correspond to the upstream region (including source B, boxes \#1 to \#9), and positive physical distances correspond to the downstream region (boxes \#11 to \#20).  
Assuming the DSA mechanism \citep{blandford+eichler87}, the relic's surface brightness sharply increases due to particle acceleration from the shock motion towards the cluster outskirts and, subsequently, slowly declines due to the inverse Compton and synchrotron energy losses, assuming that no additional acceleration mechanisms are present, that is \citep{hoeft+bruggen07,kang+17,jones+23}:
\begin{equation}\label{eq:downstream}
\Delta I(\varv_{\rm shock},\nu,B) \propto \varv_{\rm shock} \left [ \nu_{\rm obs} (1+z) \right ] ^{-1/2} \, \left (\frac{B^{1/2}}{B^2 + B_{\rm CMB}^2} \right ) \, , 
\end{equation}
where $B_{\rm CMB}=3.25(1+z)^2~\mu{\rm G}$ is the magnetic field associated with the cosmic microwave background (CMB), $B$ is the magnetic fields in the downstream region, and $\Delta I$ the relic's intrinsic width. This decline is therefore frequency-dependent and is governed by the strength of the collision (i.e. the shock Mach number $\mathcal{M}=\varv_{\rm shock}/c$, where $\varv_{\rm shock}$ is the shock velocity and $c$ is the sound speed in the ICM environment) and by the level and shape of the underlying magnetic field distribution. 
To compare our data with this picture, we adopted a log-normal distribution of the magnetic fields, $B$, of the form \citep{rajpurohit+18}:
\begin{equation}
f(B)=\frac{1}{\sqrt{2\pi}\sigma B}\exp{\left ( -\frac{\ln(B/B_0)}{2\sigma^2} \right )} \, ,
\end{equation}
where the magnetic field distribution described with a mean magnetic field strength, $B_0$, of $1~\mu$G and a log-normal width $\sigma=1$. Moreover, we assumed a Mach number of $\mathcal{M}=2.3$ \citep[calculated from the integrated radio profiles and in agreement with previous multi-frequency analysis, see][]{digennaro+23} and an opening angle of $2\Psi=30^\circ$. Following \cite{lusetti+25}, to our model we also add surface fluctuations with a Gaussian distribution of 10 kpc to account for an inhomogeneous and wiggled shock surface due to density and temperature fluctuations in the upstream region. 

We find that the data are well described by our toy model (Fig.~\ref{fig:profiles}, central panel). We note that the peaks of the two profiles are not coincident, with an offset of about 11 kpc. This shift is consistent with an intrinsic profile that widens at lower frequencies, as expected from an edge-on shock front with a downstream width determined by CR cooling, as also observed in the ridge of the ``Toothbrush'' relic \citep{rajpurohit+18}. Interestingly, we find that the 3.0 GHz observations are also well represented by a Gaussian model (dot-dashed lines) with a Full Width Half Maximum (FWHM) that equals the beam size, which suggests that the relic width is actually unresolved at this resolution and frequency. 
This does not apply for the 145 MHz data, where the FWHM of the Gaussian model is about four times larger than the beam FWHM. This model better traces the upstream emission at this frequency, while it fails to reproduce the peak and the downstream profile of the relic (see Sect. \ref{sec:upstream}).

We also estimate the spectral index profile across the relic, using Eq. \ref{eq:analyticalpha} (Fig.~\ref{fig:profiles} right panel). From box \#10 to \#13, which cover the relic's downstream region at both frequencies, we obtain a steepening from $\alpha^{145}_{2997}=-1.16\pm0.04$ to $\alpha^{145}_{2997}=-1.76\pm0.04$, consistent with the theoretical spectral index from our semi-analytical model (see black line in Fig.~\ref{fig:profiles}, central panel). Based on this value in the outermost region of the relic, we find a disagreement with the typical relationship between the injected and volume-integrated spectral indices reported in Tab.~\ref{tab:fluxes}, that is $\alpha_{\rm int}=-1.25\pm0.04$ \citep[i.e. $\alpha_{\rm inj}= \alpha_{\rm int}+0.5$,][]{kardashev62}. This could suggest a decreasing amount of injection flux of CRs electrons while it gradually slows down over time \citep{kang15}.
In all the other sectors, we are limited by the non-detection in the VLA S-band observation, and therefore we can only provide upper limits. These suggest steep spectral index values, with $\alpha<-1.4$ from boxes \#16-18. The upper limits of $\alpha<-1$ possibly reflect hints of the presence of the radio halo \citep{digennaro+23}.

\subsubsection{Shock surface}\label{sec:shock}
\begin{figure}
\centering
\includegraphics[width=0.5\textwidth]{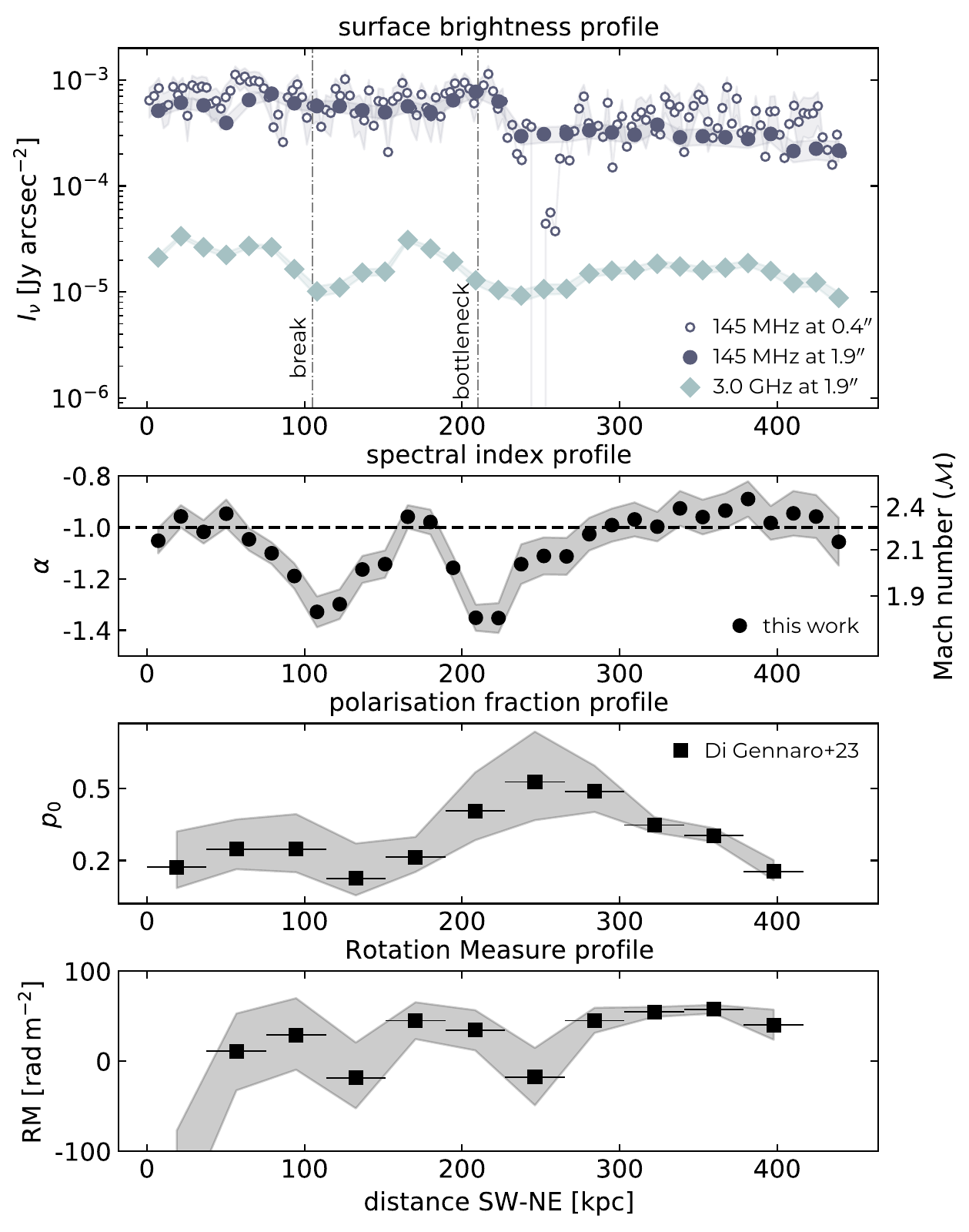}
\vspace{-6mm}
\caption{Profile along the radio shock in R2. From the top to the bottom panel: surface brightness profiles at 145 MHz ($0.4''$ and $1.9''$ resolutions, empty and filled circles, respectively) and 3.0 GHz ($1.9''$ resolution, filled diamonds), with vertical dot-dashed lines placing the position of the ``break'' and ``bottleneck''; spectral index profile at $1.9''$ between 145 MHz and 3.0 GHz, with the horizontal dashed line showing the mean value; intrinsic polarisation fraction; Rotation Measure, without the correction for the Galactic foreground \cite[$\rm RM_{Gal}=60\pm12~rad\,m^{-2}$;][]{hutschenreuter+22}. The bottom two panels are made with data taken from \cite{digennaro+23}, in the 2--4 GHz band at $5''$ resolution (see horizontal error bars). In all panels, the shaded areas define the uncertainties of each measurement.}
\label{fig:shockprofile}
\end{figure}

We investigated the shock surface (see Fig.~\ref{fig:shockprofile}) by extracting the surface brightness profiles from beam-size circles along R2, at both $0.4''$ and $1.9''$ resolutions (see Appendix \ref{apx:allimages}, Fig.~\ref{fig:images2arcsec}). The 145 MHz profiles show similar features at both resolutions, with a flat radio intensity of about $\sim8\times10^{-4} ~{\rm Jy\,arcsec^{-2}}$ up to 250 kpc, after which a sharp decrease to $\sim3\times10^{-4} ~{\rm Jy\,arcsec^{-2}}$ for the remaining relic. This location corresponds to the position of the ``bottleneck'' (see Fig.~\ref{fig:ilt-images-high} and vertical line in Fig.~\ref{fig:shockprofile}), and defines the two different parts of R2 visible in polarisation \citep[i.e., $\rm R2_N$ and $\rm R2_S$][]{digennaro+23}. This transition appears to be smoother in the 3.0 GHz data.
Contrary to these high-frequency observations, no hint of the ``break'' at $\sim100$ kpc \citep{digennaro+23} is visible in the ILT data (see vertical line in Fig.~\ref{fig:shockprofile}). This is probably due to the fact that at 145 MHz we also have the contribution of the radio bridge which is missing at GHz frequencies (see Fig.~\ref{fig:profiles} and Appendix \ref{apx:allimages}, Fig.~\ref{fig:images2arcsec} and \ref{fig:images3arcsec}).

These different features at the two frequencies result in a non-uniform distribution of the spectral index at the shock (see second topper panel in Fig.~\ref{fig:shockprofile}), where we notice two peaks of steep spectral index, in correspondence of the ``break'' and of the ``bottleneck'' ($\alpha^{145}_{2997}=-1.30\pm0.05$ and $\alpha^{145}_{2997}=-1.39\pm0.05$, respectively). Excluding these regions, the mean spectral index of the shock surface is $-1.05\pm0.02$, corresponding to a Mach number\footnote{The Mach number is calculated as  $\sqrt{\frac{2\alpha_{\rm inj}-3}{2\alpha_{\rm inj}+1}}$, where $\alpha_{\rm inj}$, namely the injected spectral index, is the spectral index at the outermost edge of the shock. Uncertainties on the Mach number are obtained through error propagation.} of $\mathcal{M}=2.15\pm0.03$. This Mach number agrees with that found from {\it Chandra} X-ray analysis \citep[$\mathcal{M}\sim1.9$, see][]{digennaro+23}. 
We compare our new high-resolution surface brightness and spectral index profiles with the published polarisation properties of the relic, obtained from the 2--4 GHz observations presented in \cite{digennaro+23}, at $5''$ resolution. The values presented in Fig.~\ref{fig:shockprofile} were taken from beam-size circles following the same path as for the $0.4''$ and $1.9''$ images. Interestingly, we note that the location of the two regions of steep spectral index also coincides with two regions of higher absolute Rotation Measure values, while an increase of the intrinsic polarisation fraction is only seen at the location of the ``bottleneck''. 
The Rotation Measure is defined as the integral along the line of sight of the combination of the parallel (i.e. along the line of sight) component of the magnetic fields ($B_\parallel$) and the thermal electron density column ($n_e$), in the form of ${\rm RM}\propto\int n_e B_\parallel\, dl$. In this formalism, the sign of the Rotation Measure simply defines the orientation of the magnetic field vector\footnote{Positive/negative RM values indicate magnetic field directed toward/away from the observer}. In the simple approach, where radio relics are located at the cluster outskirts perpendicularly to the merger axis of a binary collision, $n_e$ can be considered constant at the relic's location and, therefore, the RM value can be used to infer the magnetic field strength. At the same time, the deviation from the theoretical maximum intrinsic polarisation fraction given by the purely synchrotron ageing mechanism \citep{ensslin+11,carilli+taylor02} is related to a different contribution between the random (i.e. isotropic, $B_{\rm rand}$) and ordered (i.e. aligned to the shock surface, $B_{\rm ord}$) component of the magnetic field structure, according to $p_0\propto\alpha\,(B_{\rm rand}/B_{\rm ord})^{-2}$ \citep[see][]{sokoloff+98,govoni+04,hoeft+22}.

\begin{figure*}
\centering
\includegraphics[width=0.45\textwidth]{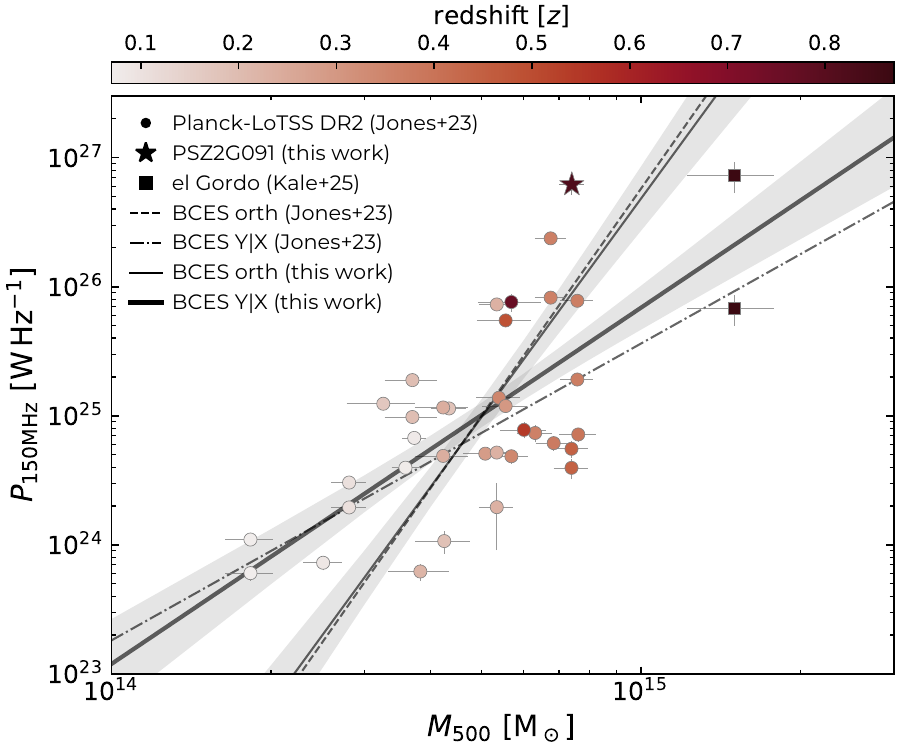}
\includegraphics[width=0.45\textwidth]{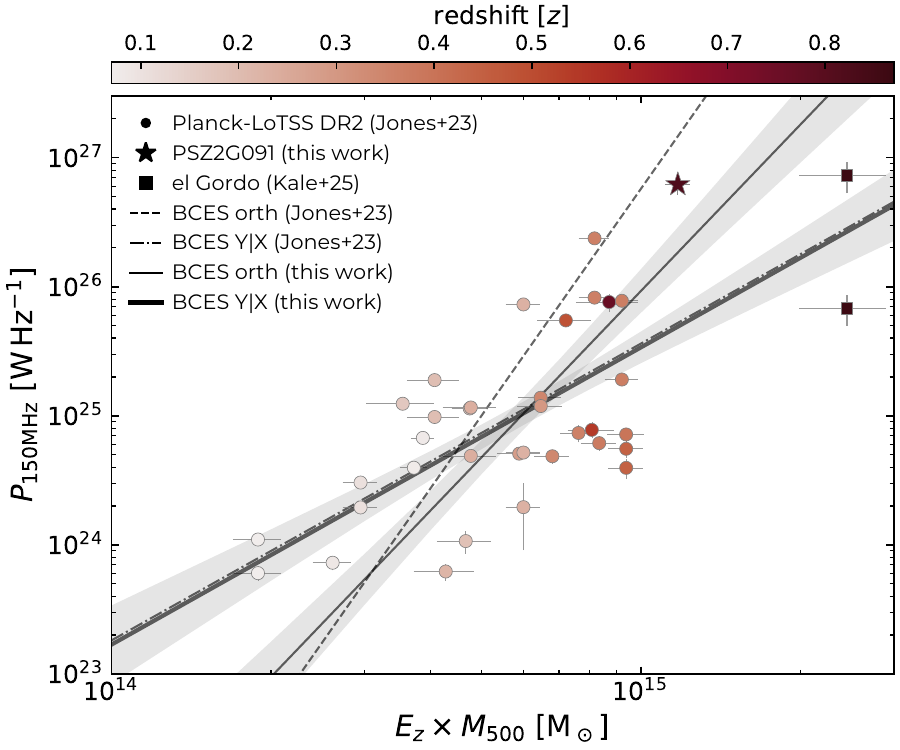}
\vspace{-3mm}
\caption{150 MHz radio power versus cluster mass diagram colour-coded based on the cluster redshift with and without the redshift evolution of the correlation (right and left panel, respectively).
Circles are from the {\it Planck} clusters with relics in LoTSS-DR2 \citep{jones+23}, without distinction of single, double or candidate radio relics; 
squares are the NW and SE relics in the ``el Gordo'' cluster re-scaled to 150 MHz using the measured spectral indices \citep[i.e. $\alpha=-1.4\pm0.3$ and $\alpha=-1.3\pm0.3$, respectively, see][]{kale+25}; 
the star is the sum of R1 and R2 in \cluster. 
Best-fit correlations are displayed as black lines: from \cite{jones+23}, the {\tt BCES orthogonal} is in light dashed and the {\tt BCES Y|X} is in light dot-dashed; from this work, which includes also \cluster\ and ``el Gordo'', the {\tt BCES orthogonal} is in light solid and the {\tt BCES Y|X} is in tick solid.
Shaded regions represent the 68\% confidence area of the correlations.}
\label{fig:radiopower_all}
\end{figure*}

\begin{table*}[]
\caption{Radio power vs cluster mass best-fit relation for radio relics.}
\vspace{-5mm}
\begin{center}
\begin{tabular}{cccccl}
\hline
\hline
Fit method & correlation & $B$ & $A$ & $\sigma_{\rm raw}$ & Reference \\
\hline
\multirow{3}{*}{\tt BCES Y|X} & $P_{\rm 150MHz}-M_{500}$ & $2.63\pm0.43$ & $-13.63\pm6.43$ & 0.58 & \multirow{2}{*}{this work} \\
 & $P_{\rm 150MHz}-E_zM_{500}$ & $2.23\pm0.34$ & $-7.92\pm5.19$ &  0.55 \\
 & $P_{\rm 150MHz}-M_{500}$& $2.30\pm0.45$ & $-8.94\pm6.59$ & & \cite{jones+23}\\
 \hline
\multirow{3}{*}{\tt BCES orthogonal} & $P_{\rm 150MHz}-M_{500}$ & $5.72\pm1.29$ & $-59.19\pm19.15$ & 0.87 & \multirow{2}{*}{this work} \\
 & $P_{\rm 150MHz}-E_zM_{500}$ & $4.34\pm0.86)$ & $-39.21\pm12.91$ & 0.75 \\
 & $P_{\rm 150MHz}-M_{500}$ & $5.84\pm1.32$ & $-60.84\pm19.39$ & & \cite{jones+23} \\
\hline
\end{tabular}%}
\end{center}
\vspace{-5mm}
\label{tab:PM-fit_all}
\end{table*}

Despite the different angular resolutions, the comparison of the spectral and polarisation profiles suggests that the ratio $B_{\rm rand}/B_{\rm ord}$ is higher at the location of the ``break'', meaning that to keep the polarisation fraction constant with a steeper spectral index and higher values of RM we need a less ordered magnetic fields. This could be due, for instance, to the contribution of the bridge coming from source B on the relic, although it is not clear whether these two kinds of emission are indeed connected. One way to verify this would be to investigate the spectral curvature profile or the spectral tomography of these sources \citep{rajpurohit+21}. However, the non-detection of both source B and the bridge at multiple higher frequencies (e.g. in the VLA L- and S-band observations) limits this analysis. On the other hand, the increase in the polarisation fraction (of a factor $\sim2$) and the change in the spectral index (of a factor $\sim1.5$) at the location of the ``bottleneck'' suggests a more ordered magnetic field. We therefore suggest a possible variation of magnetic fields along the shock front, as suggested by cosmological simulations \citep{wittor+19}.

\subsection{Radio power vs mass relation}
In the scenario where the radio relics are formed as consequence of a merger event, and where the energy released during the merger event is directly proportional to the product of the cluster masses, it is reasonable to expect a direct relation between the cluster mass and the cluster radio power. 
This is well-established for radio halos \citep[e.g.][]{cassano+13, cuciti+23, balboni+25}. 
For radio relics, analytical estimates and cosmological simulations suggest a positive correlation between their radio power and cluster mass \citep{hoeft+08,nuza+17,lee+24} but
observationally establishing such a correlation is more challenging \citep{degasperin+14,jones+23,stroe+25} given the smaller statistics\footnote{\cite{botteon+22} found occurrence rates of $\sim30\%$ and $\sim10\%$ for radio halos and relics, respectively, in the {\it Planck} clusters within the LoTSS-DR2 area.} and observational biases (e.g. Malmquist bias, projection effects, resolution). Due to these selection biases, it is also quite unclear how the relic radio power evolves with the redshift. To date, only three clusters are known to host (at least) a radio relic at $z>0.6$: ``el Gordo'' \citep[ACT-CL\,J0102.9-4916, $z=0.87$,][]{lindner+14,kale+25}, PSZ2\,G069.39+68.05 \citep[$z=0.762$,][]{botteon+22,jones+23}, and \cluster.

Using the integrated radio flux and integrated spectral index reported in Tab.~\ref{tab:fluxes}, we estimate the $k$-corrected 150 MHz radio power of the radio relic in \cluster. We find a total radio power of $P_{\rm 150MHz}=(6.2\pm1.0)\times10^{26}~{\rm W\,Hz^{-1}}$, given the sum of the integrated fluxes from Tab.~\ref{tab:fluxes}. To compare these radio powers with relics in the literature (Fig.~\ref{fig:radiopower_all}), and to derive the correlation parameters also including high-$z$ systems\footnote{\cite{jones+23} excluded the only other cluster at $z>0.6$ that hosts a radio relic, that is PSZ2\,G069.39+68.0.}, we adopted the Bivariate Correlated Errors and intrinsic Scatter \citep[BCES\footnote{\url{https://github.com/rsnemmen/BCES}}, see][]{arkitas+96} method.
We report both {\tt BCES Y|X} and the {\tt BCES orthogonal} methods applied to the data in the literature \citep{jones+23} and including the high-redshift relics. The former fitting method assumes that the  variable {\tt Y} (i.e. $P_{\rm 150MHz}$) depends on the {\tt X} one (i.e. $M_{500}$), which is in line with the assumption that relics originate from mergers. The latter method minimises the orthogonal distances between the two variables, which are assumed to be independent of each other. 
In our analysis, we assumed that $\log(P_{\rm 150MHz})=B\log(M_{500}) + A$, where $B$ is the slope and $A$ the intercept of the best-fit line. We compare our results with the best-fit reported in \cite{jones+23}, including candidate radio relics (see Tab.~\ref{tab:PM-fit_all}). We find that our best-fit lines agree within the errorbars with those reported in \cite{jones+23}. This is not surprising because we are only adding three clusters to the sample. Interestingly, we note that if we consider the redshift-evolution of the correlation (Fig.~\ref{fig:radiopower_all}, right panel), that is $\log(P_{\rm 150MHz})=B\log(E_z\,M_{500}) + A$, where $E_z=[\Omega_{\rm m,0}(1+z)^3+\Omega_\Lambda]^{1/2}$ and $\Omega_{\rm m,0}=0.3$ and $\Omega_\Lambda=0.7$ assuming a $\Lambda$CDM cosmology \citep[][]{balboni+25}, the slopes of our correlation flatten and, in the case of the {\tt BCES Y|X} method, the results are consistent with those from the low-$z$ case (i.e. with no $E_z$ correction). Even taking into account the redshift-evolution of the correlation, the slope differs from that expected by simulations \citep[$\sim1.5$ and $\sim2$ from TNG simulations, before and after corrections for the Malmquist bias respectively, see][]{lee+24}. However, we note that the slope from our {\tt BCES Y|X} fit approaches that from the TNG simulations after the correction of the Malmquist bias.
This suggests that the evolution of such a correlation exists, although it is crucial to have more high-redshift relics.

\subsection{Hints of radio emission in the shock upstream region}\label{sec:upstream}
The new ILT observations of \cluster\ have revealed the presence of diffuse emission ahead of the relic. At $1.9''$ resolution, this seems to connect solely source B and the relic (bridge, see Figs. \ref{fig:ilt-images-low} and \ref{fig:profiles}), while at $3''$ resolution it extends for a longer region, reaching a length of $\sim250$ kpc. At lower resolutions (i.e. above $3.8''\times2.7''$), corresponding to a physical scale of $\sim25$ kpc at the redshift of the cluster, these two kinds of emission blend together and cannot really be distinguished. In fact, this was not detected in previously published LOFAR images \cite{digennaro+21a,digennaro+21c}. From boxes \#7-8 in Fig.~\ref{fig:profiles}, we estimate a flux density of $\sim10$ mJy at 145 MHz and an upper limit of 0.09 mJy at 3.0 GHz. These values lead to an upper limit on the spectral index of $\alpha^{145}_{2997}<-1.6$.

The nature of the bridge and all the upstream radio emission is quite unclear. Few recent studies have shown the presence of diffuse radio emission ahead of the northern relic in the ``Sausage'' cluster \citep[R5 in CIZAJ2242.8+5301, see][]{digennaro+18,raja+24,lusetti+25}, which is believed to be a ``precursor'' shock \citep{bruggen+12}. However, R5 in the Sausage cluster is much longer ($\sim700$ kpc) and its spectral index is much flatter (i.e. $\alpha\sim-0.8$) than what we observe in \cluster. The non-detection of diffuse emission at 3.0 GHz at resolutions comparable to the ILT data suggests that this piece of emission has a steep spectrum (i.e. $\alpha<-1.6$, see Fig.~\ref{fig:profiles} right panel), meaning that we are likely detecting very old plasma. In the future, possible long-baseline LOFAR LBA observations \citep[with a resolution of $\sim1''$ at $\sim55$ MHz, see][]{wucknitz10} could help determine a more accurate integrated spectral index value of this upstream radio emission. However, such studies are currently limited to bright radio galaxies \citep{morabito+16,groeneveld+22}.
Finally, we note that, contrary to R5 in the Sausage cluster, we do not detect any polarisation signal. However, these observations are limited to the VLA $\sim5''$ resolution, where the upstream and downstream emission are blended together \citep{digennaro+23}. 
The steep spectrum and the lack of polarised emission go against the classification as a ``precursor'' shock.

Despite the challenges in classifying this kind of radio emission, this finding points to the importance of having multi-resolution (from sub to tens arcsecond) low-frequency ($\sim150$ MHz) observations to be able to detect low-surface brightness sources and disentangle different kinds of emission that otherwise would have been lost in the GHz-frequency data in high-redshift clusters.

\section{Summary and conclusion}\label{sec:conclusion}
In this work, we present for the first time a sub-arcsecond resolution study of a radio relic at frequencies below 1 GHz, in the high-redshift galaxy cluster PSZ2\,G091.83+26.11 ($z=0.822$). We used the European stations of the  LOw Frequency ARray High Band Antenna (LOFAR HBA), the so-called International LOFAR Telescope (ILT), operating in the 120--168 MHz frequency band for a total time of 16 hours. The final image of our target, that is the radio relic in the high redshift cluster PSZ2\,G091.83+26.11, reaches a resolution of $0.4''\times0.3''$ at the central frequency of 145 MHz, and a noise level of $\rms=25.4~\mu{\rm Jy\,beam}^{-1}$, which is in line with other studies at similar resolution, frequency, and integration times \citep{morabito+22,sweijen+22}. 
The achieved resolution corresponds to a spatial resolution of $\sim3$ kpc, at the redshift of the cluster. 

We found that the radio relic is resolved at sub-arcsecond resolution, although only the brightest part (i.e. R2) is visible. We performed detailed surface brightness and spectral index profiles at $1.9''$ (spatial resolution of $\sim15$ kpc), a resolution that matches the previously published VLA S-band observations \citep{digennaro+23}. We found an overall agreement with the expectation of the diffusive shock acceleration mechanism \citep[DSA,][]{blandford+eichler87}, assuming a log-normal magnetic field distribution in the downstream region. 
Interestingly, we detected radio emission in the upstream region of the shock at 145 MHz, a ``bridge'' of emission that connects the relic with a diffuse radio source. This kind of emission is undetected at high frequencies. The resulting steep spectral index ($\alpha^{145}_{2997}<-1.6$) makes this kind of emission apparently different from what is observed, for example, in the northern relic in CIZA\,J2242.8+5301 \citep{digennaro+18,raja+24,lusetti+25}, and cannot be strictly classified as a ``precursor'' shock \citep{bruggen+12}. 
Along the shock, the combination of the $1.9''$ spectral and the previous $5''$ polarisation analysis \citep{digennaro+23} suggests that there is variation of the random and ordered magnetic field ratio, with an increase of the former at the location of the ``break''. A non-negligible contribution of the bridge could explain this change of magnetic field.
Finally, we investigated the position of this high-redshift radio relic in the radio power vs cluster mass diagram at 150 MHz. We find that the relic in PSZ2\,G091.83+26.11 lies within the scatter of the correlation found for low-redshift clusters \citep{jones+23}, but the correlation flattens if we also include the redshift evolution (i.e. $P_{\rm 150MHz}-E_zM_{500}$). In this sense, we highlight the importance of detecting more high-$z$ relics, although their occurrence rate is unclear \citep{lee+24}, to define more precisely the redshift-evolution of the correlation.

In this context, thanks to the third data release of the LOFAR Two-Meter Sky Survey (LoTSS-DR3; T. Shimwell et al., submitted) that will cover 89\% of the northern sky, we will observe hundreds of high-redshift ($z>0.6$) clusters to investigate the presence of radio relics. Moreover, in the next years the International LOFAR Two-metre Sky Survey (ILoTSS) is planning to conduct a sensitive
($30~\mu\rm Jy\,beam^{-1}$) and high-resolution ($0.3''$) survey of 7,600 deg$^2$ of the extragalactic sky at $\rm DEC>20^\circ$. This survey will be co-spatial with the {\it Euclid} wide survey, allowing us to have an additional reliable high-redshift cluster catalogue where to search for and study radio relics at sub-arcsecond resolutions at $\sim150$ MHz.

\begin{acknowledgements}
%referee acknowledgements
We thank the referee for the suggestions which improved the quality of the manuscript. 
%authors acknowledgements
GDG and FdG acknowledge support from the ERC Consolidator Grant ULU 101086378. 
RT is grateful for support from the UKRI Future Leaders Fellowship (grant MR/T042842/1). This work was supported by the STFC [grants ST/T000244/1, ST/V002406/1].
MB acknowledges support from the Deutsche Forschungsgemeinschaft under Germany's Excellence Strategy - EXC 2121 ``Quantum Universe'' - 390833306 and from the BMBF ErUM-Pro grant 05A2023.
JMGHJdJ acknowledges support from project CORTEX (NWA.1160.18.316) of research programme NWA-ORC, which is (partly) financed by the Dutch Research Council (NWO), and support from the OSCARS project, which has received funding from the European Commission’s Horizon Europe Research and Innovation programme under grant agreement No. 101129751. This publication made use of computing resources from the project Deep high-resolution LOFAR imaging with file number 2023.040 of the research programme Computing Time on National Computing Facilities which is (partly) financed by the Dutch Research Council (NWO).
EDR acknowledges support by the Deutsche Forschungsgemeinschaft (DFG) and by the Fondazione ICSC, Spoke 3 Astrophysics and Cosmos Observations. National Recovery and Resilience Plan (Piano Nazionale di Ripresa e Resilienza, PNRR) Project ID CN\_00000013 “Italian Research Center for High-Performance Computing, Big Data and Quantum Computing” funded by MUR Missione 4 Componente 2 Investimento 1.4: Potenziamento strutture di ricerca e creazione di “campioni nazionali di R\&S (M4C2-19)” – Next Generation EU (NGEU).
% acknowledgements
This publication is part of the project LOFAR Data Valorization (LDV) [project numbers 2020.031, 2022.033, and 2024.047] of the research programme Computing Time on National Computer Facilities using SPIDER that is (co-)funded by the Dutch Research Council (NWO), hosted by SURF through the call for proposals of Computing Time on National Computer Facilities. 
LOFAR data products were provided by the LOFAR Surveys Key Science project (LSKSP; \url{https://lofar-surveys.org/}) and were derived from observations with the International LOFAR Telescope (ILT). LOFAR \citep{vanhaarlem+13} is the Low Frequency Array designed and constructed by ASTRON. It has observing, data processing, and data storage facilities in several countries, which are owned by various parties (each with their own funding sources), and which are collectively operated by the ILT foundation under a joint scientific policy. The ILT resources have benefited from the following recent major funding sources: CNRS-INSU, Observatoire de Paris and Université d'Orléans, France; BMBF, MIWF-NRW, MPG, Germany; Science Foundation Ireland (SFI), Department of Business, Enterprise and Innovation (DBEI), Ireland; NWO, The Netherlands; The Science and Technology Facilities Council, UK; Ministry of Science and Higher Education, Poland; The Istituto Nazionale di Astrofisica (INAF), Italy.
This research made use of the Dutch national e-infrastructure with support of the SURF Cooperative (e-infra 180169) and the LOFAR e-infra group. The J\"ulich LOFAR Long Term Archive and the German LOFAR network are both coordinated and operated by the J\"ulich Supercomputing Centre (JSC), and computing resources on the supercomputer JUWELS at JSC were provided by the Gauss Centre for Supercomputing e.V. (grant CHTB00) through the John von Neumann Institute for Computing (NIC).
This research made use of the University of Hertfordshire high-performance computing facility and the LOFAR-UK computing facility located at the University of Hertfordshire and supported by STFC [ST/P000096/1]. This research made use of the LOFAR-IT computing infrastructure supported and operated by INAF, including the resources within the PLEIADI special "LOFAR" project by USC-C of INAF, and by the Physics Dept. of Turin University (under the agreement with Consorzio Interuniversitario per la Fisica Spaziale) at the C3S Supercomputing Centre, Italy. 
%VLA acknowledgements
The National Radio Astronomy Observatory is a facility of the National Science Foundation operated under cooperative agreement by Associated Universities, Inc.
%other
This research made use of {\tt APLpy}, an open-source plotting package for Python \citep{aplpy}, {\tt astropy}, a community-developed core Python package for Astronomy \citep{astropy+13,astropy+18}, {\tt matplotlib} (Hunter 2007), {\tt numpy} \citep{harris+20}, {\tt scipy} \citep{virtanen+20}, and {\tt shadems} (\url{https://github.com/ratt-ru/shadeMS/}).
\end{acknowledgements}

\bibliographystyle{aa}
\bibliography{PAPER.bib}

\clearpage
\onecolumn
\begin{appendix}

\clearpage
\section{All frequency images}\label{apx:allimages}
In this section, we show the additional frequency images of \cluster. 

\begin{figure*}[h!]
%\centering
\sidecaption
\includegraphics[width=12cm]{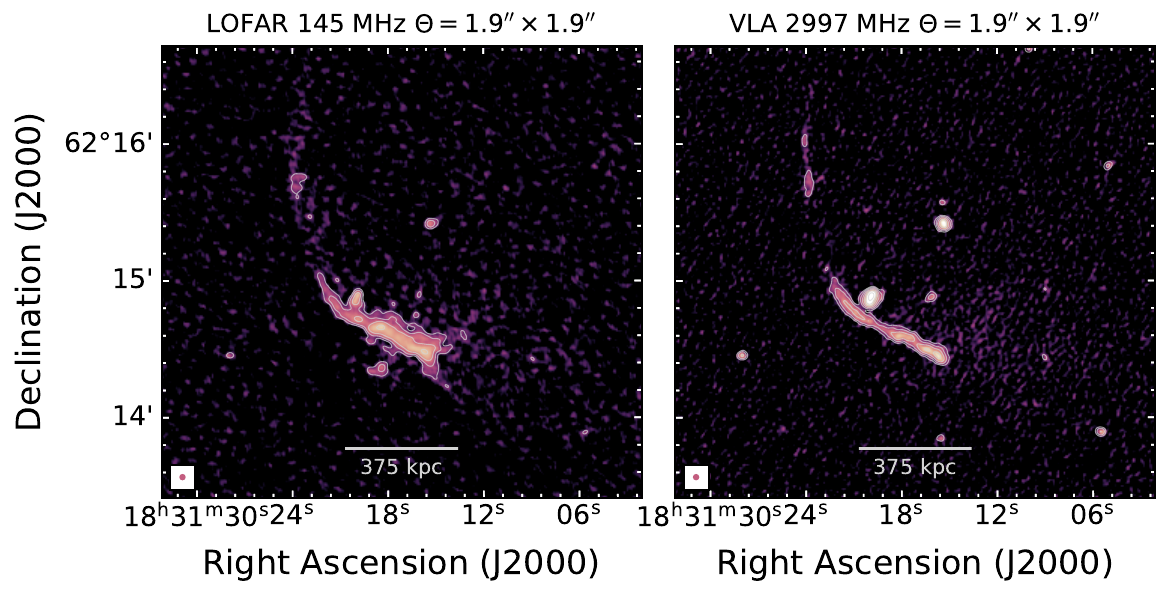} 
\caption{ILT 145 MHz and VLA 3.0 GHz at $1.9''$ resolution. These images were used to produce the spectral index map in Fig.~\ref{fig:spixmpas}, left panel.}\label{fig:images2arcsec}
\end{figure*}

\begin{figure*}[h!]
%\centering
\sidecaption
\includegraphics[width=12cm]{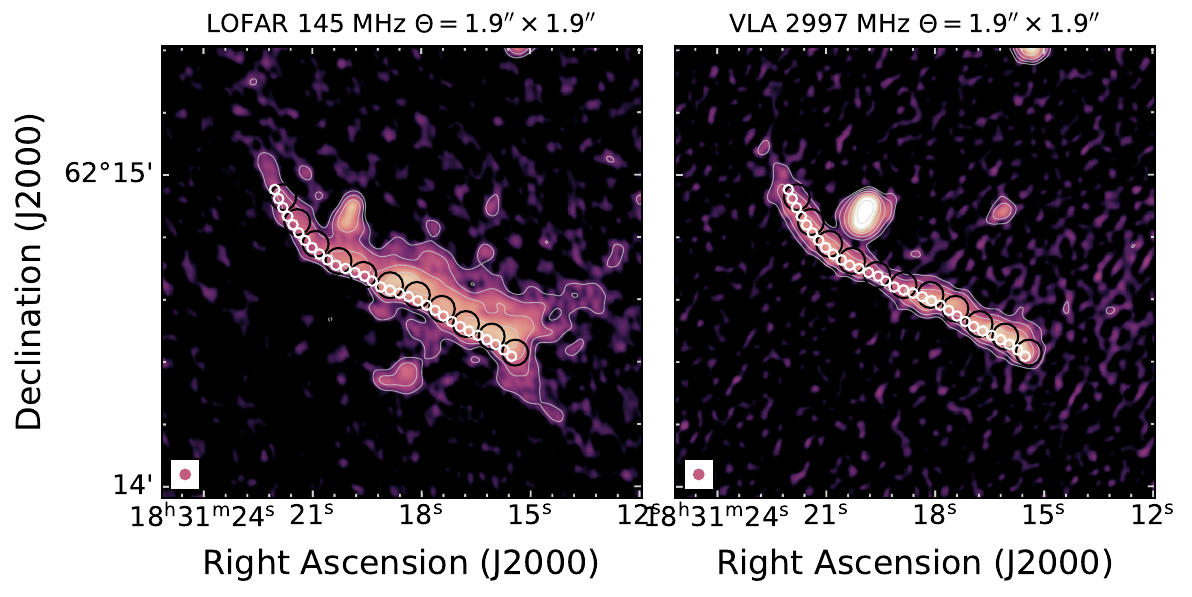} 
\caption{Zoom-in on R2 at $1.9''$ resolution (ILT 145 MHz, left, and VLA 3.0 GHz, right) with the region used to produce the profiles in Fig.~\ref{fig:shockprofile} (in white, surface brightness and spectral index profiles at $1.9''$ resolution; in black, polarisation and RM profiles at $5''$ resolution).}\label{fig:images2arcsec}
\end{figure*}

\begin{figure*}[h!]
\centering
%\sidecaption
\includegraphics[height=4cm]{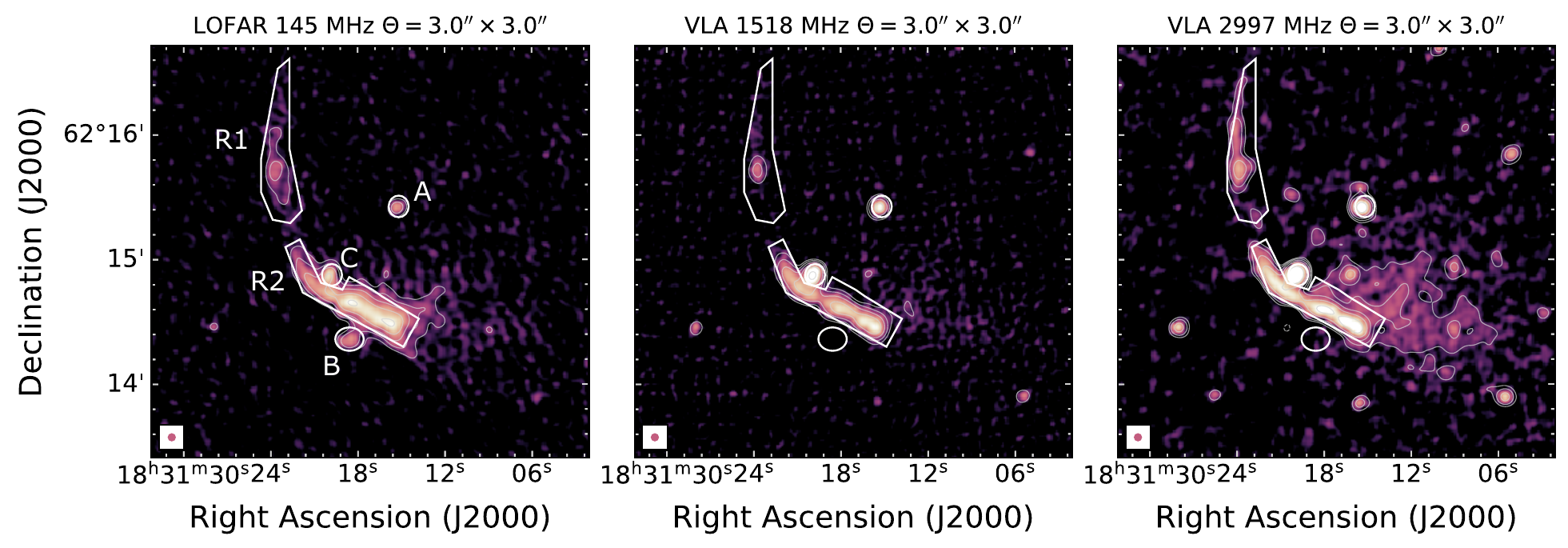} 
\includegraphics[height=4cm]{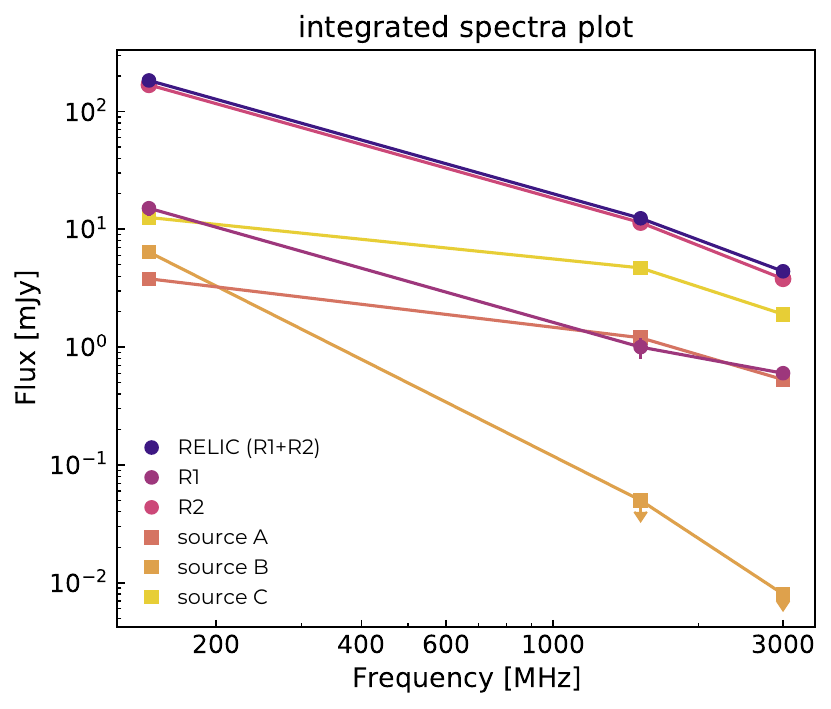}
\caption{ILT 145 MHz, VLA at 1.5 GHz and VLA 3.0 GHz at $3.0''$ resolution. These images were used to produce the spectral index and curvature maps in Fig.~\ref{fig:spixmpas}, central and right panels. White polygons show the region chosen for the flux densities reported in Tab.~\ref{tab:fluxes}.}\label{fig:images3arcsec}
\end{figure*}

\end{appendix}

\end{document}